\documentclass[twocolumn,aps,10pt,float]{revtex4-1}
\usepackage{epsfig}
\begin{document}
\title{\Large Exact general solution to the three-dimensional Ising model and a self-consistency equation for the nearest-neighbors' correlations}

\author{Vitaly V. Kocharovsky$^{1,2}$ and Vladimir V. Kocharovsky$^{2}$\\
\textit{$^{1}$Department of Physics and Astronomy, Texas A\&M University, College Station, TX 77843-4242, USA}\\
\textit{$^{2}$Institute of Applied Physics, Russian Academy of Science,
603950 Nizhny Novgorod, Russia}}

\date{\today}

\begin{abstract}
    We find an exact general solution to the three-dimensional (3D) Ising model via an exact self-consistency equation for nearest-neighbors' correlations. It is derived by means of an exact solution to the recurrence equations for partial contractions of creation and annihilation operators for constrained spin bosons in a Holstein-Primakoff representation. In particular, we calculate analytically the total irreducible self-energy, the order parameter, the correlation functions, and the joined occupation probabilities of spin bosons. The developed regular microscopic quantum-field-theory method has a potential for a full solution of a long-standing and still open problem of 3D critical phenomena.
    
\end{abstract}

\maketitle

\section{Three-dimensional Ising model in a Holstein-Primakoff representation of the constrained spin bosons}

   The three-dimensional (3D) problem of critical phenomena in the continuous phase transitions remains one of the major unsolved problems in theoretical physics. In particular, an exact solution of a famous 3D Ising model (or any other nontrivial 3D model) of a phase transition was not found, despite almost a century of intensive effort. The Ising model was invented by W. Lenz in 1920. Only the 1D \cite{Ising} and 2D Ising models with zero \cite{Onsager} or non-zero \cite{Zamolodchikov} external field were solved (see \cite{Baxter1989,Kadanoff,CritPhen-RG1992} and references therein). The Ising model is, probably, the most studied model in statistical physics  of critical phenomena, since (i) it is considered as a basis prototype of many-body systems with short-range interactions showing nontrivial critical behavior and (ii) it describes phase transitions in several important systems, such as uniaxial antiferromagnets, simple fluids with liquid-vapor  transition, multicomponent fluid mixtures, and micellar systems \cite{Baxter1989,Kadanoff,CritPhen-RG1992,Vicari2002}. 

    Recently, we found a microscopic theory of phase transitions in a critical region \cite{PLA2015,PhysicaScripta2015}, which is based on a nonperturbative method of the recurrence equations for partial operator contractions, and suggested to use it for the solution of the 3D Ising model \cite{PhysicaScripta2015,PLA2015Ising}.
  
   The present paper is a detailed account of that solution, including the exact solutions (and their derivations) for the partial two-operator contractions (Sec. II), the total irreducible self-energy (Sec. III), the joined probability distributions of spin-bosons' occupations (Sec. IV), the order parameter and the correlation functions (Sec. V) as well as the derivation of exact closed self-consistency equation for the nearest-neighbors' correlations (Sec. VI) and general formula for the permanents of a circulant matrix and its submatrices (Appendix). A current Sec. I introduces a concept of the constrained spin bosons \cite{PhysicaScripta2015,PLA2015Ising} which allows us to upgrade a Holstein-Primakoff representation \cite{HolsteinPrimakoff} to a rigorous theory on a constrained many-body Hilbert space.
   
    Of course, the presented results constitute only the first, basis steps towards a full solution of the 3D Ising model. A lot of further work is required for a computation of various statistical and  thermodynamic quantities, critical functions and exponents, etc. on the basis of the proposed exact  solution.
   
    A beauty of the found exact solution is in its remarkably transparent, universal algebraic structure (see, for example, Eqs. (\ref{newSigma(l)}), (\ref{rho1per}), (\ref{2x2qcumulants}), (\ref{spinIsing}), (\ref{g'Ising}), (\ref{magnetization-per}) below). It allows us to employ all power of algebra and matrix calculus, including a famous MacMahon master theorem on an inverse determinant, its novel generalization for an inverse matrix (Eqs. (\ref{MMM}), (\ref{KMT2})), related multivariate cumulant analysis, theory of circulant Toeplitz matrices, etc., as well as the determinants, pfaffians, permanents (Eq. (\ref{per})) and block-matrix permanents (Eqs. (\ref{matrixper}), (\ref{2x2qcumulants})) of certain matrices. That canonical structure clearly suggests its regular generalization for the exact solution to other major models of phase transitions.  

   We consider a 3D cubic lattice of $N$ interacting quantized spins $s= \frac{1}{2}$ with a period $a$ in a box with volume $L^3$ and periodic boundary conditions. However, the method is valid for an arbitrary dimensionality of lattice $d=1, 2, 3, ...$. The lattice sites are enumerated by a position vector ${\bf r}$. According to the Holstein-Primakoff representation \cite{HolsteinPrimakoff}, worked out also by Schwinger \cite{Schwinger1965}, each spin is a system of two spin bosons, which are constrained to have a fixed total occupation 
\begin{equation}
\hat{n}_{0{\bf r}}+\hat{n}_{\bf r} =2s; \quad \hat{n}_{\bf r}=\hat{a}^{\dagger}_{\bf r} \hat{a}_{\bf r}, \quad \hat{n}_{0{\bf r}}=\hat{a}^{\dagger}_{0{\bf r}} \hat{a}_{0{\bf r}}.
\label{bosons}
\end{equation}
 The $\hat{a}_{\bf r}$ and $\hat{a}_{0{\bf r}}$ are the annihilation operators obeying the Bose canonical commutation relations: $[\hat{a}_{\bf r}, \hat{a}_{\bf r'}^{\dagger }]=\delta_{{\bf r},{\bf r'}}$, $[\hat{a}_{0{\bf r}}, \hat{a}_{0{\bf r'}}^{\dagger }]=\delta_{{\bf r},{\bf r'}}$, and all $({\bf r})$-operators commute with all $(0{\bf r'})$-operators; $\delta_{{\bf r},{\bf r'}}$ is a Kronecker $\delta$-function. A vector spin operator $\hat{\bf S}_{\bf r}$ at a site ${\bf r}$ is given by its components:
\begin{equation}
\hat{S}^x_{\bf r}=\frac{\hat{a}^{\dagger}_{0{\bf r}}\hat{a}_{\bf r} + \hat{a}^{\dagger}_{\bf r}\hat{a}_{0{\bf r}}}{2}, \ \hat{S}^y_{\bf r}=\frac{\hat{a}^{\dagger}_{0{\bf r}}\hat{a}_{\bf r}- \hat{a}^{\dagger}_{\bf r}\hat{a}_{0{\bf r}}}{2i}, \ \hat{S}^z_{\bf r}= s - \hat{a}^{\dagger}_{\bf r}\hat{a}_{\bf r}.
\label{vectorspin}
\end{equation}

   A proper reduction of a many-body Hilbert space ensures \cite{PLA2015} that this system is isomorphic to a system of $N$ spin-boson excitations, described by annihilation operators $\hat{\beta}_{\bf r}$ at each site ${\bf r}$ and obeying the Bose canonical commutation relations $[\hat{\beta}_{\bf r}, \hat{\beta}_{\bf r'}^{\dagger }]=\delta_{{\bf r},{\bf r'}}$, if we cutoff them by a step-function $\theta(2s-\hat{n}_{\bf r})$; $\theta(x)=1$ if $x\geq 0$ and $\theta(x)=0$ if $x<0$. This isomorphism is valid on an entire physically allowed Hilbert space and is achieved by equating the annihilation operators $\hat{\beta}^{'}_{\bf r}=\hat{\beta}_{\bf r}\theta(2s-\hat{n}_{\bf r})$ of those constrained, true excitations to the cutoff Holstein-Primakoff's transition operators: 
\begin{equation}
\hat{\beta}^{'}_{\bf r}=\hat{a}_{0{\bf r}}^{\dagger}(1+2s-\hat{n}_{\bf r})^{-1/2}\hat{a}_{\bf r}\theta(2s-\hat{n}_{\bf r}).
\label{spinalpha}
\end{equation}
Here and thereinafter we add a prime to a symbol of an unconstrained quantity to denote its cutoff, constrained counterpart. The vector components of spin operator are
\begin{equation}
\hat{S}^x_{\bf r}=\frac{1}{2} ({S}^{-}_{\bf r}+\hat{S}^{+}_{\bf r}), \ \hat{S}^y_{\bf r}=\frac{i}{2} ({S}^{-}_{\bf r}-\hat{S}^{+}_{\bf r}), \ \hat{S}^z_{\bf r}= s- \hat{n}_{\bf r},
\label{spincomponents}
\end{equation}
where spin raising and lowering operators are equal to
\begin{equation}
\hat{S}^{+}_{\bf r} = \sqrt{2s-\hat{n}_{\bf r}} \hat{\beta}_{\bf r}^{'}, \hat{S}^{-}_{\bf r}= \hat{\beta}_{\bf r}^{'\dagger}\sqrt{2s-\hat{n}_{\bf r}}; \ \hat{n}_{\bf r}= \hat{\beta}_{\bf r}^{'\dagger}\hat{\beta}_{\bf r}^{'}.
\label{spinraise}
\end{equation}
The aforementioned isomorphism is not trivial since it is not valid outside the constrained, physically allowed Hilbert space and the commutation relations for the creation and annihilation operators of the true spin excitations in Eqs. (\ref{spinalpha}) are not canonical,
\begin{equation}
\qquad [\hat{\beta}_{\bf r}^{'}, \hat{\beta}_{\bf r'}^{'\dagger}]=\delta_{{\bf r},{\bf r'}}(1- (2s+1) \delta_{\hat{n}_{\bf r},2s}) .
\label{spinnonCCR}
\end{equation}

        A free Hamiltonian of a system of $N$ spins in a lattice
\begin{equation}
H_0 = \sum_{\bf r} \varepsilon \hat{n}_{\bf r}, \quad \hat{n}_{\bf r}=\hat{\beta}_{\bf r}^{\dagger}\hat{\beta}_{\bf r}, \quad \varepsilon= g \mu_B B_{ext},
\label{spinH_0}
\end{equation}
is determined by a Zeeman energy $-g\mu_B B_{ext}\hat{S}^z$ of a spin in an external magnetic field $B_{ext}$ (which is assumed homogeneous and directed along the axis $z$) via a $g$-factor and a Bohr magneton $\mu_B =\frac{e\hbar}{2Mc}$. We intentionally define the free Hamiltonian in Eq. (\ref{spinH_0}) via the unconstrained occupation operators $\hat{n}_{\bf r}=\hat{\beta}_{\bf r}^{\dagger}\hat{\beta}_{\bf r}$ on a full Fock space generated by a set of creation operators $\{\hat{\beta}_{\bf r}^{\dagger}\}$, that is on the extended many-body Hilbert space without any $\theta(2s-\hat{n}_{\bf r})$ cutoff factors. That makes the free Hamiltonian purely quadratic which is necessary for a validity of the standard diagram technique. The latter is crucial for a derivation of the Dyson-type equations, like Eq. (\ref{spinGEq}). One is allowed to skip the $\theta(2s-\hat{n}_{\bf r})$ cutoff factors in $H_0$ in virtue of an equality $\hat{\beta}_{\bf r}^{\dagger}\hat{\beta}_{\bf r}= \hat{\beta}_{\bf r}^{'\dagger}\hat{\beta}_{\bf r}^{'}$, valid on the physical many-body Hilbert space, and a fact that the occupation operator $\hat{n}_{\bf r}=\hat{\beta}_{\bf r}^{\dagger}\hat{\beta}_{\bf r}$ leaves that space invariant.
   
   An interaction Hamiltonian of the Ising model 
\begin{equation}
H'= -\sum_{\bf r} \sum_{{\bf r'}\neq {\bf r}} J_{{\bf r},{\bf r'}}\hat{S}_{\bf r}^z\hat{S}_{\bf r'}^z 
\label{HxyHz},
\end{equation}
in view of the isomorphism's Eqs. (\ref{spinalpha})-(\ref{spinraise}), takes a form 
\begin{equation}
H' = -\sum_{{\bf r}\neq {\bf r'}} J_{{\bf r},{\bf r'}} [s-\theta (2s-\hat{n}_{\bf r})\hat{n}_{\bf r}] [s-\theta (2s-\hat{n}_{\bf r'})\hat{n}_{\bf r'}] .
\label{HIsing}
\end{equation}
Here a coupling between spins is a symmetric function $J_{{\bf r},{\bf r'}}= J_{{\bf r}-{\bf r'}}$ of a vector ${\bf r}-{\bf r'}$, connecting spins. For a spin at a site ${\bf r_0}$ there are only the coordination number $p$ of the nonzero couplings $J_{{\bf r_0},{\bf r}_l} \neq 0$ with the neighboring spins at sites ${\bf r}_l= {\bf r_0} + {\bf l}; l=1, ..., p$. The result in Eq. (\ref{HIsing}) generalizes the Holstein-Primakoff's one \cite{HolsteinPrimakoff} by including the nonpolynomial operator $\theta(2s-\hat{n}_{\bf r})$-cutoff functions, which add a spin-constraint nonlinear interaction and are crucially important in a critical region. 

    Since Holstein-Primakoff's paper of 1940, there were many previous attempts to convert it into a rigorous and tractable microscopic theory of critical phenomena in magnetic phase transitions, e.g., \cite{Dembinski1964,Tyablikov1967}. No one was successful. For example, a Dyson's theory of spin waves in a ferromagnet \cite{Dyson1956} is invalid in the critical region and restricts an analysis to just a well-formed ordered phase. Due to a lack of a proper mathematical apparatus, first of all, the partial contraction of operators and diagram technique for the nonpolynomial averages, Dyson thought that "the Holstein-Primakoff formalism is thus essentially nonlinear and unamenable to exact calculations".

   A total Hamiltonian $H=H_0 +H^{'}$ defines, for any operator $\hat{A}$, a Matsubara operator $\tilde{A}_{\tau}=$ $e^{\tau H}\hat{A}e^{-\tau H}$ evolving in an imaginary time $\tau \in [0, \frac{1}{T}]$ in a Heisenberg representation. A symbol $T$ denotes a temperature. A symbol $\tilde{A}_{j\tau}$ stands for an operator itself $\tilde{A}_{1\tau} =\tilde{A}_{\tau}$ at $j=1$ and a Matsubara-conjugated operator $\tilde{A}_{2\tau} =\tilde{\bar{A}}_{\tau}$ at $j=2$. 

    The unconstrained and true Matsubara Green's functions for spin excitations are defined by a $T_{\tau}$-ordering:
\begin{equation}
G^{j_2\tau_2{\bf r_2}}_{j_1\tau_1{\bf r_1}}= -\langle T_{\tau} \tilde{\beta}_{j_1\tau_1{\bf r_1}}\tilde{\bar{\beta}}_{j_2\tau_2{\bf r_2}} \rangle , 
\label{spinGreen}
\end{equation}
\begin{equation}
G^{'j_2\tau_2{\bf r_2}}_{j_1\tau_1{\bf r_1}}= -\langle T_{\tau} \tilde{\beta'}_{j_1\tau_1{\bf r_1}}\tilde{\bar{\beta'}}_{j_2\tau_2{\bf r_2}} \hat{\theta} \rangle /P_s ; \quad P_s = \langle \hat{\theta} \rangle .
\label{spinGreen'}
\end{equation}
Here an unconstrained thermal average over an equilibrium statistical operator $\rho = e^{-\frac{H}{T}}/\text{Tr}\{ e^{-\frac{H}{T}} \}$ of spin-boson excitations is denoted by angles as
\begin{equation}
\langle \dots \rangle \equiv  \text{Tr}\{\dots e^{-\frac{H}{T}}\}/\text{Tr}\{e^{-\frac{H}{T}}\} 
\label{average}
\end{equation}
and a true, constrained thermal average is denoted as $\langle \dots \hat{\theta} \rangle /P_s$. A normalization factor $P_s= \langle \hat{\theta} \rangle $ is equal to a cumulative probability of all occupations of spin excitations in the unconstrained Fock space to be within physically allowed intervals $n_{\bf r} \in [0,2s]$ for all lattice sites ${\bf r}$; $\hat{\theta} =  \prod_{\bf r} \theta(2s-\hat{n}_{\bf r})$ is a product of all $N$ cutoff factors.  

  In the Ising model there is no coherence, $\langle \beta_{{\bf r}\tau} \rangle =0$, and the unconstrained Green's functions obey the usual Dyson equation with a total irreducible self-energy $\Sigma^{j_2x_2}_{j_1x_1}$,
\begin{equation}
(G_{j_1x_1}^{j_2x_2}) = (G_{j_1x_1}^{(0)j_2x_2}) + \check{G}^{(0)}[\check{\Sigma}[G_{j_1x_1}^{j_2x_2}]]. 
\label{spinGEq}
\end{equation}
Here the integral operators $\check{\Sigma}$ or $\check{G}^{(0)}$, applied to any function $f_{jx}$ of an index $j$ and a four-dimensional coordinate $x=\{\tau,{\bf r}\}$, stand for a convolution of that function $f_{jx}$ over the variables $j, \tau, {\bf r}$ with the total irreducible self-energy $\Sigma$ or the free propagator $G^{(0)}$, respectively:
\begin{equation}
\check{K}[f_{jx}]\equiv \sum_{j'=1}^2 \sum_{\bf r'} \int_0^{1/T} K^{j'x'}_{jx} f_{j'x'} d \tau' \ \text{for} \ \check{K}=\check{\Sigma}, \check{G}^{(0)};
\label{convolution}
\end{equation}
\begin{equation}
G_{j_1x_1}^{(0)j_2x_2} = - \frac{\delta_{j_1,j_2}\delta_{{\bf r_1},{\bf r_2}}}{e^{(-1)^{j_1}\varepsilon (\tau_2-\tau_1)}} \Big[ \frac{1}{e^{\varepsilon /T}-1}+\theta [(-1)^{j_1}(\tau_2-\tau_1)] \Big] .
\label{spinG0}
\end{equation}
  
    The total irreducible self-energy is defined by equation
\begin{equation}
\langle T_{\tau} [\tilde{\beta}_{j_1x_1}, \tilde{H}^{'}_{\tau_1}]\tilde{\bar{\beta}}_{j_2x_2} \rangle =(-1)^{j_1}\sum_{j=1}^2 \int_0^{\frac{1}{T}} \sum_{\bf r}\Sigma_{j_1x_1}^{jx}G_{jx}^{j_2x_2} d\tau .
\label{spinself-energy}
\end{equation}
We find an operator in the left hand side of Eq. (\ref{spinself-energy}) as
\begin{equation}
[\tilde{\beta}_{\tau_1{\bf r_1}}, \tilde{H}^{'}_{\tau_1}]\tilde{\bar{\beta}}_{\tau_2{\bf r_2}} = \sum_{{\bf r}\neq {\bf r_1}} J_{{\bf r},{\bf r_1}} \tilde{\beta}_{\tau_1{\bf r_1}} f(\tilde{n}_{\bf r_1},\tilde{n}_{\bf r}) \tilde{\bar{\beta}}_{\tau_2{\bf r_2}} 
\label{commutator}
\end{equation}
and use a fact that for the Ising model in Eqs. (\ref{spinH_0})-(\ref{HIsing}) the Matsubara occupation operator $\tilde{n}_{\tau{\bf r}}=\tilde{n}_{\bf r}$ does not depend on imaginary time $\tau$. In the case of spins $s= \frac{1}{2}$, an operator-valued function $f$ consists of two components:
$$f(\tilde{n}_{\bf r_1},\tilde{n}_{\bf r})= f^{(1)} + f^{(2)}, \quad f^{(1)}(\tilde{n}_{\bf r_1})= \delta_{1,\tilde{n}_{\bf r_1}}-\delta_{2,\tilde{n}_{\bf r_1}},$$
\begin{equation}
f^{(2)}(\tilde{n}_{\bf r_1},\tilde{n}_{\bf r}) = 2(\delta_{2,\tilde{n}_{\bf r_1}}-\delta_{1,\tilde{n}_{\bf r_1}}) \delta_{1,\tilde{n}_{\bf r}}.
\label{f}
\end{equation}
The first one $f^{(1)}(\tilde{n}_{\bf r_1})$ depends only on one occupation operator $\tilde{n}_{\bf r_1}$, while the second function $f^{(2)}(\tilde{n}_{\bf r_1},\tilde{n}_{\bf r})$ depends on two occupation operators $\tilde{n}_{\bf r_1}$ and $\tilde{n}_{\bf r}$.

\section{Exact general solution to the recurrence equations for the partial two-operator contractions}

    In order to calculate the nonpolynomial averages, like the ones for the self-energy in Eq. (\ref{spinself-energy}) and the true Green's functions in Eq. (\ref{spinGreen'}), we employ the recurrence equations for partial operator contractions, derived via a nonpolynomial diagram technique in \cite{PLA2015}. The point is that the constrained, true Green's functions do not obey equations of a Dyson type due to a presence of the nonpolynomial functions $\theta(2s-\hat{n}_{\bf r})$ and a standard diagram technique is not suited to deal with them. In particular, we express the true Green's functions in a form
\begin{equation}
G^{'J_2}_{J_1}= -\langle \tilde{b}^{J_2}_{J_1}[\tilde{\theta}_{\tau_1} \tilde{\theta}_{\tau_2}] \rangle /P_s ,
\label{spinG'}
\end{equation}
which includes a basis partial two-operator contraction
\begin{equation}
\tilde{b}_{J_1}^{J_2}[f(\{\tilde{n}_{x'_1},\tilde{n}_{x'_2}\})] \equiv \mathcal{A}_{\tau_{i_1}\tau_{i_2}} T_{\tau} \{ \tilde{\beta}_{J_1}^{c} \tilde{\bar{\beta}}_{J_2}^{c} f^c(\{\tilde{n}_{x'_1},\tilde{n}_{x'_2}\}) \}.
\label{spinb}
\end{equation}

The latter is an operator-valued functional, evaluated for an operator function $f$ and defined as a sum of all possible partial connected contractions, denoted by superscripts "c". Let us consider a generic case of an arbitrary operator function $f(\{\tilde{n}_{x'_1},\tilde{n}_{x'_2}\} )$, which depends on two sets $ \{\tilde{n}_{\tau'_1{\bf r'_1}}\}$, $ \{\tilde{n}_{\tau'_2{\bf r'_2}}\}$ of spin-excitation occupation operators at lattice sites $\{{\bf r'_1}\}, \{{\bf r'_2}\}$ and times $\tau'_1$, $\tau'_2$. An anti-normal ordering $\mathcal{A}_{\tau_{i_1}\tau_{i_2}}$ prescribes only positions of the external operators $\tilde{\beta}_{J_1}$ and $\tilde{\bar{\beta}}_{J_2}$ relative to the function $f( \{\tilde{n}_{x'_1},\tilde{n}_{x'_2}\})$ and does not affect any other operators' positions, set by $T_{\tau}$-ordering. We use the short notations for the combined indexes $J= \{ji{\bf r_i}\}$ and $J_l= \{j_l i_l{\bf r_{i_l}}\}$. An index $i=1,2$ (or $i_l$) enumerates different times $\tau_i$ (or $\tau_{i_l}$) in the external operator $\tilde{\beta}_{j\tau_i{\bf r_i}}$ (or $\tilde{\beta}_{j_l\tau_{i_l}{\bf r_{i_l}}}$).

    The exact closed recurrence (difference) equations for the basis partial operator contraction $\tilde{b}^{J_2}_{J_1}[f]$ for an arbitrary function $f(\{ m_{J'} \}) =f(\{\tilde{n}_{x'_1}+2s+1-m_{x'_1}, \tilde{n}_{x'_2}+2s+1-m_{x'_2}\})$, where a set $\{ m_{J'}\}$ consists of two sets  of integers $\{m_{x'_1}\}$ and $\{m_{x'_2}\}$, are derived in \cite{PLA2015}:
\begin{equation}
\tilde{b}^{J_2}_{J_1}[f] =g_{J_1}^{J'_1} \Delta_{m_{J'_1}} \Delta_{m_{J'_2}} \tilde{b}^{J'_2}_{J'_1}[f]g_{J'_2}^{J_2} - g_{J_1}^{J'} \Delta_{m_{J'}}f g_{J'}^{J_2}  -g_{J_1}^{J_2}f.
\label{spin2-contraction}
\end{equation}
Here a matrix $g_J^{J'}$ is the unconstrained Green's function $G_J^{J'}$ for $\tau_i \neq\tau_{i'}$ and its limit at $\tau_i \to \tau_{i'} -(-1)^{j'}\times 0$ for equal times in accord with an anti-normal ordering of operators $\tilde{\beta}_J$, $\tilde{\bar{\beta}}_{J'}$. The latter is dictated by the anti-normal ordering in the definition of the basis contractions in Eq. (\ref{spinb}). In Eq. (\ref{spin2-contraction}), a symbol $\Delta_{m_{J'}}$ means a partial difference operator \cite{PDE-Cheng,DE-Agarwal,DE-Elaydi} ($\Delta_{m_1} f({m_1,m_2})=f({m_1+1,m_2})-f({m_1,m_2})$ and $\Delta_{m_2} f({m_1,m_2})=f({m_1,m_2+1})-f({m_1,m_2})$), and we assume an Einstein's summation over the repeated indexes $J', J'_1, J'_2$. The sums run over $j'=1,2$ and all different arguments $\tilde{n}_{x'_{i'}}$ of $f$ for $J'$ and similarly for $J'_1, J'_2$. 

   A linear system (\ref{spin2-contraction}) of the integral equations over the spin positions' variables and discrete (recurrence) equations over variables $\{m_{J'}\}$ can be solved by well-known methods \cite{PDE-Cheng,DE-Agarwal,DE-Elaydi}, such as a Z-transform, a characteristic function, or a direct recursion. The partial contraction in Eq. (\ref{spinG'}) is given by those solutions at $m_{J'}=2s+1$. 

   For the Ising model, due to independence of $\tilde{n}_{\tau{\bf r}}=\tilde{n}_{\bf r}$ on $\tau$, it is enough to consider the aforementioned limit of equal times and, hence, to use only the reduced combined indexes $I=\{j,{\bf r}\}$ and $I'=\{j',{\bf r'}\}$, that is to skip the indexes $i, i'$ of times $\tau_i, \tau_{i'}$. The corresponding $2N\times 2N$-matrices, say $\tilde{b}_{j{\bf r}}^{j'{\bf r'}}$, consist of pairs of rows and columns. Each pair is enumerated by the site-position index ${\bf r}$ or ${\bf r'}$, while the creation-annihilation index $j=1,2$ or $j'=1,2$ enumerates two rows or two columns in each pair, respectively. We find the most powerful, and convenient for further applications, general solution of recurrence equations (\ref{spin2-contraction}) in a form of an inverse $2N\times 2N$-matrix,
\begin{equation}   
\tilde{b} [f] = -Z(Z+g)^{-1}g  [f] \equiv -(1+gZ^{-1})^{-1}g  [f] .
\label{b-solution}
\end{equation} 
It can be applied to an arbitrary function-argument $f$. One can prove the solution (\ref{b-solution}) by direct substitution in Eq. (\ref{spin2-contraction}). By definition, a matrix $Z$ acts on $f$ as operator and has nonzero elements only at its main diagonal:
\begin{equation}
Z= \text{diag}\{Z_I^I \}\equiv \{ \frac{\delta_{{\bf r},{\bf r'}}\delta_{j,j'}}{1-D_{\bf r}}\}, \ Z_I^I = Z_{\bf r} \equiv \frac{1}{1-D_{\bf r}} . 
\label{Z}
\end{equation}
Those elements do not depend on the creation-annihilation index $j$, but only on the site-position index ${\bf r}$, and are equal to inverse operators $(1-D_{\bf r})^{-1}$, where a lowering operator $D_{\bf r}$ decreases an argument $\tilde{n}_{\bf r}$ of function $f$ by unity and leaves all other arguments $\tilde{n}_{\bf r'}$ intact, 
\begin{equation}
D_{\bf r}[f(\{ \tilde{n}_{\bf r'} \})] = f(\tilde{n}_{\bf r}-1,\{ \tilde{n}_{\bf r'}, {\bf r'}\neq {\bf r} \}). 
\label{D}
\end{equation}
An inverse matrix operator $(1-D)^{-1}$ can be found via a geometrical progression over the raising operators $d_{\bf r} = D_{\bf r}^{-1}, d_{\bf r}[f(\{ \tilde{n}_{\bf r'} \})] = f(\tilde{n}_{\bf r}+1,\{ \tilde{n}_{\bf r'}, {\bf r'}\neq {\bf r} \})$, as follows
\begin{equation}
(1-D)^{-1}= -d(1+d+d^2+...), \qquad d\equiv D^{-1},
\label{inverse(1-D)}
\end{equation}
if that sum converges for a given function $f(\{\tilde{n}_{\bf r} \})$. 

    Especially important and simple is a particular case when the function-argument $f(\{\tilde{n}_{\bf r_k}\})$ depends only on a finite number $m$ of occupation operators $\tilde{n}_{\bf r_k}$ at lattice sites ${\bf r_k}, k=1,2,...,m$. Then, for all other lattice sites ${\bf r'}\neq {\bf r_k}$ one has a trivial operator $D_{\bf r'} \equiv 1$, so that the corresponding factors $gZ_{\bf r'}^{-1}$ in Eq. (\ref{b-solution}) vanish, $Z_{\bf r'}^{-1} \equiv 1-D_{\bf r'}=0$, leaving only a trivial unity diagonal 
\begin{equation}
\Big(\frac{1}{1+gZ^{-1}}\Big)_{j'\bf r'}^{j''\bf r''}=\delta_{j',j''}\delta_{{\bf r'},{\bf r''}} \text{for} \ {\bf r'},{\bf r''}\neq {\bf r_k}, k=1,...,m,
\label{2mx2m}
\end{equation}
in the matrix $(1+gZ^{-1})^{-1}$ in Eq. (\ref{b-solution}). In this case the matrix $Z(Z+g)^{-1} \equiv (1+gZ^{-1})^{-1}$ in the general solution (\ref{b-solution}) should be treated as a $2m\times 2m$-block for the lattice sites ${\bf r_k}, k=1,2,...,m,$ and as a unity matrix for all other sites. The geometrical progression expansion (\ref{inverse(1-D)}) should be used only for that $2m\times 2m$-block.

\section{Exact solution for the total irreducible self-energy}

    We consider a homogeneous phase, when the Green's function $G^{j_2\tau_2{\bf r_2}}_{j_1\tau_1{\bf r_1}}$ depends on ${\bf r_1}$ and ${\bf r_2}$ only via ${\bf r_2}-{\bf r_1}$. So, it is a Toeplitz matrix with respect to indexes ${\bf r_1}$, ${\bf r_2}$. The exact total irreducible self-energy has a form
\begin{equation}
\Sigma_{J_0}^{J} =-\delta(\tau-\tau_0) \sum_{l=1}^{p} \sum_{I'} J_{{\bf r_0},{\bf r_l}} \bar{b}_{I_0}^{I'}[f(\tilde{n}_{\bf r_0},\tilde{n}_{\bf r_l})] (g^{-1})_{I'}^{I},  
\label{IsingSelfEnergy}
\end{equation}
that follows from Eqs. (\ref{spinself-energy})-(\ref{f}) and contains an averaged two-operator contraction $\bar{b}_{I_0}^{I'}[f]=\langle\tilde{b}_{I_0}^{I'}[f]\rangle$. The latter can be calculated by means of the exact solution in Eq. (\ref{b-solution}). A product of the equal-time anti-normally ordered correlation matrix $g$ from that solution and its inverse matrix $(g^{-1})_{I'}^I$ in Eq. (\ref{IsingSelfEnergy}) makes unity, and we get a remarkably simple, exact solution for the self-energy
\begin{equation}
\Sigma_{J_0}^{J} =\delta(\tau-\tau_0) \sum_{l=1}^{p} J_{{\bf r_0},{\bf r_l}} \langle Z\frac{1}{Z+g} [f(\tilde{n}_{\bf r_0},\tilde{n}_{\bf r_l})] \rangle^{I}_{I_0} .
\label{exactSelfEnergy}
\end{equation}

    According to Eq. (\ref{f}), the function $f$ in Eqs. (\ref{IsingSelfEnergy}), (\ref{exactSelfEnergy}) has two components, $f(\tilde{n}_{\bf r_0},\tilde{n}_{\bf r_l})=f^{(1)}(\tilde{n}_{\bf r_0}) +f^{(2)}(\tilde{n}_{\bf r_0},\tilde{n}_{\bf r_l})$. Following the argument in Eq. (\ref{2mx2m}), we conclude that they contribute to the self-energy matrix with the $2\times 2$- and $4\times 4$-blocks, respectively. Hence, the self-energy matrix is a diagonal, $2(p+1)$-banded in indexes $I_0=\{j_0,{\bf r_0}\}$ and $I=\{j,{\bf r}\}$, matrix 
\begin{equation}
\Sigma_{J_0}^{J} =\delta(\tau-\tau_0) \sum_{l=1}^{p} J_{{\bf r_0},{\bf r_l}} \bar{\Sigma}_{I_0}^{I}(l) , \quad {\bf r_l}= {\bf r_0} + {\bf l},
\label{pSelfEnergyIsing}
\end{equation}
in which, for a given site ${\bf r_0}$ in a lattice, each $4\times 4$-block
\begin{equation}
(\bar{\Sigma}_{I_0}^{I})(l)= \Big(\frac{\rho_1}{S} +\frac{\rho_0}{S^2}\Big)\otimes E 
\label{newSigma(l)}
\end{equation}
$$-2\rho_{1,1}q^{-1}-2\Big[\rho_{0,1}q^{-1}E_1q^{-1}+\rho_{1,0}q^{-1}E_2q^{-1}\Big]$$
$$+2\rho_{0,0}\Big[q^{-1}E_1q^{-1}E_2q^{-1}+q^{-1}E_2q^{-1}E_1q^{-1}\Big]$$
originates from correlations with nearest spin-neighbor $l=1, ...,p$, has nonzero elements only for the corresponding pair of the nearest-neighbors' positions ${\bf r}={\bf r_0}, {\bf r_0} + {\bf l}$, and contains only a related correlation $4\times4$-matrix
\begin{equation} 
q_I^{I'}(l) \equiv g_{j{\bf R}}^{j'{\bf R'}} =-\langle \mathcal{A} \hat{\beta}_{j{\bf R}} \hat{\beta}^{\dagger}_{j'{\bf R'}} \rangle , \ q(l)= \Big( \frac{S \ \ | \ C}{C^{\dagger}| \ S} \Big).
\label{q4x4}
\end{equation}
The $q=q^{\dagger}$ is hermitian, the ${\bf R}$ and ${\bf R'}$ run over two values $\{ {\bf r_0}, {\bf r_l} \}$, $\mathcal{A}$ means anti-normal ordering, $2\times2$-matrices $g_{j}^{j'}(l)= g_{j{\bf r_0}}^{j'{\bf r_l}}$ of basis auto- and cross-correlations are denoted as $g(0)=S=S^{\dagger}$ and $g(l\neq 0)=C(l)$, respectively. In Eq. (\ref{newSigma(l)}), $\otimes$ means a tensor product of $2\times2$-matrices, the $4\times4$-matrices $E_1$, $E_2$ are defined as block matrices:
\begin{equation}     
E_1 = \Big( \frac{E | {\bf 0}}{\ {\bf 0}| {\bf 0}} \Big), \quad E_2 = \Big( \frac{{\bf 0} | {\bf 0}}{\ {\bf 0}| E} \Big), 
\label{E}
\end{equation}    
where the $E$ and $\bf{0}$ are the unity and zero $2\times2$-matrices.   

    This result is obtained from Eq. (\ref{exactSelfEnergy}) by means of the geometrical-progression expansion (\ref{inverse(1-D)}) and the fact that only the following, very few lower-order combinations of operators $Z_{\bf r_0}$ and $Z_{\bf r_l}$, Eqs. (\ref{Z})-(\ref{inverse(1-D)}), produce nonzero averages from the functions $f^{(1)}$ and $f^{(2)}$ in Eq. (\ref{f}):
\begin{equation} 
\langle Z_{\bf r_0}[f^{(1)}(\tilde{n}_{\bf r_0})] \rangle = \rho_1 , \quad \langle Z^2_{\bf r_0}[f^{(1)}(\tilde{n}_{\bf r_0})] \rangle = -\rho_0 ,
\label{rho0,rho1}
\end{equation}
\begin{equation}
\langle Z_{\bf r_0}[f^{(2)}(\tilde{n}_{\bf r_0},\tilde{n}_{\bf r_l})] \rangle =-2\rho_{1,1}, \ \langle Z^2_{\bf r_0}[f^{(2)}] \rangle = 2\rho_{0,1} ,
\label{rho00,rho10,rho11}
\end{equation}
$$\langle Z_{\bf r_l}[f^{(2)}(\tilde{n}_{\bf r_0},\tilde{n}_{\bf r_l})] \rangle = 2(\rho_{1,0}- \rho_{2,0}),$$
$$\langle Z_{\bf r_0}Z_{\bf r_l}[f^{(2)}(\tilde{n}_{\bf r_0},\tilde{n}_{\bf r_l})] \rangle = 2\rho_{1,0} , \ \langle Z^2_{\bf r_0}Z_{\bf r_l}[f^{(2)}] \rangle = 2\rho_{0,0}.$$
These equations and self-energy $4\times4$-blocks in Eq. (\ref{newSigma(l)}) contain the non-cutoff probabilities $\rho_{n_{\bf r_0}}=\langle \delta_{\tilde{n}_{\bf r_0}, n_{\bf r_0}} \rangle$ and $\rho_{n_{\bf r_0},n_{\bf r_l}}=\langle \delta_{\tilde{n}_{\bf r_0}, n_{\bf r_0}} \delta_{\tilde{n}_{\bf r_l}, n_{\bf r_l}} \rangle$ for the spin bosons at sites ${\bf r_0}$ and ${\bf r_l}$ to acquire the $n_{\bf r_0}$ and $n_{\bf r_l}$ quanta of excitations.

    All terms in Eq. (\ref{newSigma(l)}) come from the corresponding to Eqs. (\ref{rho0,rho1}), (\ref{rho00,rho10,rho11}) terms in a Taylor expansion of the matrix $Z\frac{1}{Z+g}[f]$ in Eq. (\ref{exactSelfEnergy}) over variables $Z_{\bf r_0}$ and $Z_{\bf r_l}$. They were calculated by means of the well-known formulas for partial derivatives of a matrix, inverse to $B=Z+g$:
\begin{equation}    
\frac{\partial B^{-1}}{\partial Z_{\bf r_0}}= -\frac{1}{B} \frac{\partial B}{\partial Z_{\bf r_0}} \frac{1}{B} , \quad \frac{\partial B^{-1}}{\partial Z_{\bf r_l}}= -\frac{1}{B} \frac{\partial B}{\partial Z_{\bf r_l}} \frac{1}{B} , 
\label{derivative}
\end{equation}
$$\frac{\partial^{2} B^{-1}}{\partial Z_{\bf r_0}\partial Z_{\bf r_l}}= \frac{1}{B} \frac{\partial B}{\partial Z_{\bf r_0}}\frac{1}{B} \frac{\partial B}{\partial Z_{\bf r_l}} \frac{1}{B} + \frac{1}{B}\frac{\partial B}{\partial Z_{\bf r_l}} \frac{1}{B}\frac{\partial B}{\partial Z_{\bf r_0}}\frac{1}{B} .$$
The first term in the self-energy (\ref{newSigma(l)}) originates from the component $f^{(1)}$ in the commutator (\ref{commutator}), while all the rest terms - from the component $f^{(2)}$ (see Eq. (\ref{f})).

    In general, the correlation $4\times4$-matrix $q(l)$ and the joined probabilities of spin bosons' occupations $\rho_{n_{\bf r_0},n_{\bf r_l}}$ in the right hand side of Eq. (\ref{newSigma(l)}) for the self-energy $4\times 4$-blocks $\bar{\Sigma}(l)$ depend on a particular position of the nearest neighbor ${\bf l}={\bf r_l} -{\bf r_0}$, although, for a sake of brevity, that dependence is not shown explicitly.

    The self-energy in Eqs. (\ref{pSelfEnergyIsing}), (\ref{newSigma(l)}) can be rewritten explicitly in terms of the introduced in Eq. (\ref{q4x4}) $2\times2$-matrices of basis auto- and cross-correlations as follows
\begin{equation}
\Sigma_{J_0}^{J} =\delta(\tau-\tau_0) \sum_{l=0}^{p} \delta_{{\bf r},{\bf r_l}} \Sigma_{j_0{\bf r_0}}^{j{\bf r_l}}(l) , \quad {\bf r_l}= {\bf r_0} + {\bf l},
\label{p+1SelfEnergyIsing}
\end{equation}
where the $2\times2$-matrix blocks $\Sigma(l)=(\Sigma_{j_0{\bf r_0}}^{j{\bf r_l}}(l))$, of course, are different from $4\times4$-blocks $\bar{\Sigma}_{I_0}^{I}(l)$ in Eqs. (\ref{pSelfEnergyIsing}), (\ref{newSigma(l)}):
\begin{equation}
\Sigma(0)= \sum_{l=1}^{p} J_{{\bf r_0},{\bf r_l}} [\rho_1 S^{-1} +\rho_0 S^{-2} - 2\rho_{1,1}K-2\rho_{0,1}K^2
\label{Sigma(0)}
\end{equation}
$$-2\rho_{1,0}KCS^{-2}C^{\dagger}K+2\rho_{0,0}K(KCS^{-2}C^{\dagger}+CS^{-2}C^{\dagger}K)K],$$
\begin{equation}  
\Sigma(l \neq 0)= 2J_{{\bf r_0},{\bf r_l}} [(\rho_{1,1}+\rho_{0,1}K+\rho_{0,0}KC\frac{1}{S^2}C^{\dagger}K)KC\frac{1}{S}
\label{Sigma(l)}
\end{equation} 
$$+(\rho_{1,0}-\rho_{0,0}K)KC\frac{1}{S^2}(1+C^{\dagger}KC\frac{1}{S})]; \ K=\frac{1}{S-CS^{-1}C^{\dagger}}.$$ 
The latter result coincides with Eqs. (92)-(93) of \cite{PhysicaScripta2015} and (15)-(16) of \cite{PLA2015Ising}, if one nullifies the cross-correlation matrix $C$ in all terms, which include the single-site probabilities $\rho_0$ and $\rho_1$. That was obviously implied (but was miswritten) in \cite{PhysicaScripta2015,PLA2015Ising}, since calculations of both $f^{(1)}$- and $f^{(2)}$-contributions were made by direct recursion of Eq. (\ref{spin2-contraction}) via a common $4\times 4$-block, so that the result for $f^{(1)}$-contribution should be further reduced to $2\times 2$-block, in accord with Eq. (\ref{2mx2m}), by nullifying the cross-correlation $C$. So, the calculations in the present paper and in \cite{PhysicaScripta2015,PLA2015Ising} provide a cross-check that fully confirms the result. 

    For the exact solution of the Ising model, it is crucial to get that exact result for the total irreducible self-energy, which allows one to go beyond standard second-order or ladder approximations. A canonical algebraic structure of the exact result in Eqs. (\ref{pSelfEnergyIsing}), (\ref{newSigma(l)}) clearly suggests its regular generalization for the exact solution to various other models of phase transitions.

\section{Joined unconstrained statistics of the spin-bosons' occupations}

    The next step is analysis of a joined non-cutoff distribution of spin-bosons' occupations at all $N$ lattice sites
\begin{equation}
\rho_{\{ n_{\bf r}\}} \equiv \langle \prod_{{\bf r}={\bf r_1},...,{\bf r_N}} \delta_{\tilde{n}_{\bf r},n_{\bf r}} \rangle .
\label{rhoN}
\end{equation}
Actually, for an exact solution of 3D Ising model we need to calculate its particular values for a joined probability 
\begin{equation}
\rho_{1\{m\}}\equiv \rho_{\{ n_{\bf r}=1: {\bf  r}={\bf r_1},...,{\bf  r_m}; \ n_{\bf r'}=0: {\bf  r'}\neq {\bf r_1},...,{\bf  r_m}\}} = \langle f_m \rangle,
\label{rho1m}
\end{equation}
\begin{equation}
f_m = \prod_{{\bf r}={\bf r_k}, k=1,...,m} \delta_{\tilde{n}_{\bf r},1}\prod_{{\bf r'}\neq {\bf r_k}, k=1,...,m} \delta_{\tilde{n}_{\bf r'},0} ,
\label{fm}
\end{equation}
to have unity occupations $n_{{\bf  r_k}}=1$ for $m$ spin bosons at a subset of sites $\{ m \}=\{{\bf  r_k}, k=1,...,m\}$ and zero occupations for all other $N-m$ spin bosons in the lattice, since the latter probability (\ref{rho1m}) determines the true joined statistics of spin-bosons' occupations and the true correlation functions, which we calculate in Sec. V.  

    Also, for Eq. (\ref{newSigma(l)}), we need a similar joined unconstrained distribution of spin-bosons' occupations at only a subset of lattice sites $\{ M \}=\{{\bf  r_k}, k=1,...,M\}$,
\begin{equation}
\rho^{\{M\}}_{\{ n_{\bf r}\}} \equiv \langle \prod_{{\bf r}={\bf r_1},...,{\bf r_M}} \delta_{\tilde{n}_{\bf r},n_{\bf r}} \rangle , \qquad M \leq N ,
\label{rhoM}
\end{equation}
which admits arbitrary occupations $n_{\bf r'}=0,1,2,...,\infty$, ${\bf r'}\neq {\bf r_1},...,{\bf r_M},$ at all other $N-M$ lattice sites, i.e., it is non-cutoff averaged over the latter occupations. Again, we need its particular values for a joined probability 
\begin{equation}
\rho_{1\{m\}}^{\{M\}}\equiv \rho^{\{M\}}_{\{ n_{\bf r}=1: {\bf  r}={\bf r_1},...,{\bf  r_m}; \ n_{\bf r'}=0: {\bf  r'}={\bf r_{m+1}},...,{\bf  r_M}\}} = \langle f_m^{\{M\}} \rangle,
\label{rho1mM}
\end{equation}
\begin{equation}
f_m^{\{M\}} = \prod_{{\bf  r}={\bf r_1},...,{\bf  r_m}} \delta_{\tilde{n}_{\bf r},1}\prod_{{\bf r'}={\bf r_{m+1}},...,{\bf  r_M}} \delta_{\tilde{n}_{\bf r'},0} , \ m\leq M\leq N,
\label{fmM}
\end{equation}
to have unity occupations $n_{{\bf  r_k}}=1$ for $m$ spin bosons at a subset of sites $\{ m \}=\{{\bf  r_k}, k=1,...,m\}\subseteq \{ M \}$, zero occupations for $M-m$ spin bosons at a subset of sites $\{ M \}\setminus \{ m \}=\{{\bf  r_k}, k=m+1,...,M\}\subseteq \{ M \}$, and arbitrary occupations for all other $N-M$ spin bosons. 

     We employ the corresponding characteristic functions
\begin{equation}
\Theta_N (\{ u_{\bf r} \}) = \langle \exp \Big( i\sum_{{\bf r}={\bf r_1},...,{\bf r_N}} u_{\bf r}\tilde{n}_{\bf r}\Big) \rangle ,
\label{ThetaN}
\end{equation}
\begin{equation}
\Theta^{\{M\}}_N (\{ u_{\bf r} \}) = \langle \exp \Big( i\sum_{{\bf r}={\bf r_1},...,{\bf r_M}} u_{\bf r}\tilde{n}_{\bf r}\Big) \rangle ,
\label{ThetaM}
\end{equation}
the derivatives of which yield those joined distributions:
\begin{equation}
\rho_{\{ n_{\bf r}\}} = \prod_{{\bf r}={\bf r_1},...,{\bf  r_N}} \Big( \frac{1}{n_{\bf r}!} \frac{\partial^{n_{\bf r}}}{\partial z_{\bf r}^{n_{\bf r}}} \Big) \Theta_N \Big|_{\{z_{\bf r}=0\}}, \ z_{\bf r}=e^{iu_{\bf r}},
\label{rhoNdefinition}
\end{equation}
\begin{equation}
\rho^{\{M\}}_{\{ n_{\bf r}\}} = \prod_{{\bf r}={\bf r_1},...,{\bf  r_M}} \Big( \frac{1}{n_{\bf r}!} \frac{\partial^{n_{\bf r}}}{\partial z_{\bf r}^{n_{\bf r}}} \Big) \Theta^{\{M\}}_N \Big|_{\{z_{\bf r}=0\}}. 
\label{rhoMdefinition}
\end{equation}
We find the characteristic function $\Theta_N$ from a system of obvious equations
\begin{equation}
z_{\bf r}\frac{\partial \Theta_N}{\partial z_{\bf r}} =-\Theta_N +\frac{1}{z_{\bf r}}\langle \tilde{b}_{1{\bf r}}^{1{\bf r}}[e^{i\sum_{\bf r'} u_{\bf r'}\tilde{n}_{\bf r'}}] \rangle , \ {\bf r}={\bf r_1},...,{\bf r_N},
\label{EqsThetaN}
\end{equation}
which include an average value $\langle \tilde{b}_{1{\bf r}}^{1{\bf r}}[f] \rangle$ of the partial two-operator contraction, taken for the operator function $f=e^{i\sum_{\bf r'} u_{\bf r'}\tilde{n}_{\bf r'}}$. The general solution in Eq. (\ref{b-solution}) yields 
\begin{equation}
\langle \tilde{b}_{1{\bf r}}^{1{\bf r}}[e^{i\sum_{\bf r'} u_{\bf r'}\tilde{n}_{\bf r'}}] \rangle=-((1+gZ_{\Theta}^{-1})^{-1} g)_{1{\bf r}}^{1{\bf r}} \Theta_N , 
\label{b-Theta}
\end{equation}
where a diagonal matrix $Z_{\Theta}$ is not the operator matrix in Eq. (\ref{Z}) anymore, but a function of the variables $\{z_{\bf r}=e^{iu_{\bf r}}\}$ of the characteristic function:
\begin{equation}
Z_{\Theta}= \text{diag}\{(Z_{\Theta})_I^I \}, \quad (Z_{\Theta})_I^{I'} = \frac{z_{\bf r}\delta_{{\bf r},{\bf r'}}}{z_{\bf r}-1} \delta_{j,j'}. 
\label{ZTheta}
\end{equation}
Given a $N$-dimensional gradient by Eqs. (\ref{EqsThetaN}), we solve that system of the first order partial differential equations for the $\ln \Theta_N$ by explicit integration between points $P_{i-1}=\{z_{\bf r_1},...,z_{\bf r_{i-1}},1,...,1\}$ and $P_{i}=\{z_{\bf r_1},...,z_{\bf r_i},1,...,1\}$ along each axis $z_{\bf r_i}$,
\begin{equation}
\ln \Theta_N = -\sum_{i=1}^{N} \int_{P_{i-1}}^{P_{i}} \Big[ 1+ \frac{({(1+g{Z'}_{\Theta}^{-1} )}^{-1} g)_{1{\bf r_i}}^{1{\bf r_i}}}{z'_{\bf r_i}} \Big] \frac{dz'_{\bf r_i}}{z'_{\bf r_i}} .
\label{lnTheta}
\end{equation}
A representation of a diagonal element of inverse matrix $(g+Z_{\Theta})^{-1}$ via partial derivative of the determinant $\text{det}(g+Z_{\Theta})$ with respect to variable $Z_{_{\Theta}{\bf r}}=z_{\bf r}/(z_{\bf r}-1)$,
\begin{equation}
((g+Z_{\Theta})^{-1})_{1{\bf r}}^{1{\bf r}}=\frac{1}{2\text{det}(g+Z_{\Theta})}\frac{\partial \text{det}(g+Z_{\Theta})}{\partial Z_{_{\Theta}{\bf r}}},
\label{lndet}
\end{equation}
helps to calculate the integral analytically, and we find
$$\Theta_N (\{ u_{\bf r} \}) = \frac{1}{\sqrt{\text{det}(g+Z_{\Theta})}} \prod_{{\bf r}={\bf r_1},...,{\bf  r_N}} \frac{1}{1-z_{\bf r}}$$ 
\begin{equation}
= \frac{1}{\sqrt{\text{det}g}} \frac{1}{\sqrt{\text{det}(1-(1+g^{-1})z)}}, \ z_I^{I'}=z_{\bf r}\delta_{{\bf r},{\bf r'}} \delta_{j,j'} .
\label{ThetaNresult}
\end{equation}
Here the diagonal matrices $Z_{\Theta}$ and $z$ are related as $Z_{\Theta}=z/(z-1)$. Finally, we checked that Eq. (\ref{ThetaNresult}) is a correct solution for the characteristic function by its direct substitution into Eq. (\ref{EqsThetaN}) and proving that the equation is exactly satisfied. Of course, the obtained solution in Eq. (\ref{ThetaNresult}) is normalized to unity at a point $\{ u_{\bf r} =0\}$, $\Theta_N (\{ u_{\bf r} =0\}) = 1$, as it should be for a characteristic function of any distribution. 

     The probability of unity occupations for $m$ spin bosons and zero occupations for all other spin bosons, Eq. (\ref{rho1m}), is set by differentiation of that characteristic function: 
\begin{equation}
\rho_{1\{m\}}= \frac{\partial^m \Theta_N}{\partial z_{\bf r_1}... \partial z_{\bf r_m}} \Big|_{\{z_{\bf r}=0\}} .
\label{rho1mTheta}
\end{equation}     
It is a coefficients in front of multilinear term $z_{\bf r_1}...z_{\bf r_m}$ in a Taylor expansion of the characteristic function $\Theta_N$ over the variables $\{ z_{\bf r} \}$ at the zero point $\{ z_{\bf r}=0 \}$. 

    One way to evaluate that Taylor expansion is to employ a well-known MacMahon master theorem \cite{Percus1971}, that yields a Taylor expansion of a function, inversely proportional to a determinant of matrix $1-Ax$, over variables $\{x_i\}$,
\begin{equation}
\frac{1}{\text{det}(1-Ax)}= \sum_{s_1,...,s_N} \text{per}^{(s_1,...,s_N)}A \ x_1^{s_1}... x_N^{s_N} ,
\label{MMM}
\end{equation} 
where $s_i \geq 0$ is a non-negative integer ($i=1,...,N$), $x=\text{diag}\{ x_1, ..., x_N \}$ a diagonal matrix, $A$ a $N\times N$-matrix, and $\text{per}^{(s_1,...,s_N)}A$ a generalized permanent of matrix $A$. For the required by Eq. (\ref{rho1mTheta}) multilinear terms with a subset of unity integers $\{s_i=1; i=1,...,m\}$ and the rest of integers being equal zero, the corresponding permanents are reduced to the standard permanents,
$$\text{per}^{(\{s_i=1; i=1,...,m\},\{s_j=0; j=m+1,...,N\})}A$$
\begin{equation}
= \text{per} A_{\{ m \}} = \sum_{\sigma} \prod_{i=1}^m A_{i}^{\sigma(i)} ,
\label{per}
\end{equation}
which are defined by the same sum of the products of elements of the corresponding $m\times m$-matrix $A_{\{ m \}}$ as the one in $\text{det}A_{\{ m \}}=\sum_{\sigma} \text{sgn}(\sigma ) \prod_{i=1}^m A_{i}^{\sigma(i)}$ with all sign factors $\text{sgn}(\sigma )$ replaced by $+1$, i.e., the signatures $\text{sgn}(\sigma )$ of the permutations $\sigma$ of $(1,...,m)$ are not taken into account. Then, in order to get the derivatives in Eq. (\ref{rho1mTheta}), one may compute a multilinear expansion of the characteristic function in Eq. (\ref{ThetaNresult}) by taking into account (i) an additional to MacMahon master Eq. (\ref{MMM}) square-root function via the corresponding Bell polynomials of that Fa$\acute{\text{a}}$ di Bruno's formula and (ii) an equality of the variables $z_{1{\bf r}}^{1{\bf r}}=z_{2{\bf r}}^{2{\bf r}}=z_{\bf r}$ in two adjacent columns with the same site-index ${\bf r}$. Especially simple is a case of vanishing anomalous correlations, $g_{1{\bf r}}^{2{\bf r'}}=0$ for $\forall {\bf r}, {\bf r'}$, and non-zero normal correlations $g_{1{\bf r}}^{1{\bf r'}} =g_{2{\bf r}}^{2{\bf r'}}$, which are real-valued in the homogeneous phases, when the matrix $g_{j{\bf r}}^{j'{\bf r'}}$ is a circulant Toeplitz matrix in indexes ${\bf r}, {\bf r'}$ and the arbitrary phases of spin-bosons' annihilation operators $\hat{\beta}_{\bf r}$ are calibrated properly. In this case, one has $\text{det}[g+Z_{\Theta}] =[\text{det}[(g+Z_{\Theta})_{1{\bf r}}^{1{\bf r'}}]]^2$ and Eqs. (\ref{ThetaNresult})-(\ref{per}) yield 
\begin{equation}
\Theta_N = \frac{1}{\text{det}[g_1-(1+g_1)z_1]}, \ g_1\equiv (g_{1{\bf r}}^{1{\bf r'}}), \ z_1\equiv (z_{\bf r}\delta_{{\bf r},{\bf r'}}),
\label{rho1per}
\end{equation}
$$\rho_{1\{m\}}= \frac{\text{per} A_{\{m\}}^{(1)}}{\text{det}g_1}; \ A_{\{m\}}^{(1)} \equiv [(1+g_1^{-1})_{\bf r}^{\bf r'}], {\bf r}, {\bf r'}={\bf r_1},...,{\bf r_m}.$$
Here the elements of the $N\times N$-matrices $g_1$ and $z_1$ as well as $m\times m$-matrix $(1+g_1^{-1})_{\{m\}}$ are labeled solely by the site-indexes ${\bf r}, {\bf r'}$. The effect of normal cross-correlations $g_{1{\bf r}}^{1{\bf r'}}\neq 0$ between spin bosons at different lattice sites on their joined unconstrained statistics, described by Eq. (\ref{rho1per}), remains highly nontrivial even in that case of vanishing anomalous correlations $g_{1{\bf r}}^{2{\bf r'}}=0$ for $\forall {\bf r}, {\bf r'}$. 

    A related technique could be based on a well-known in algebra notion of a pfaffian of a $2N\times 2N$-matrix, which is equal to a square root of the matrix's determinant, say, $\text{pf} (g+Z_{\Theta}) =\sqrt{\text{det}(g+Z_{\Theta})}$ for the matrix $g+Z_{\Theta}$ in Eq. (\ref{ThetaNresult}). However, the method of pfaffians implies the skew-symmetric matrices (see, for example, \cite{pfaffian2009,Kitaev2001,Wimmer2012} and references therein), that is only a special case of the hermitian correlation matrix $g$, corresponding to the pure imaginary off-diagonal elements.  

    Another way to evaluate the Taylor expansion of the characteristic function $\Theta_N$, related to Eq. (\ref{rho1mTheta}), is to follow a more canonical, cumulant analysis. Namely, one starts with evaluation of the Taylor expansion of the logarithm of characteristic function, for example,
\begin{equation}
\ln \Theta_N = \sum_{\{ {\bf r_i}, i=i_1,...,i_m \}} \sum_{s_{i_1},...,s_{i_m}} \frac{\kappa'_{\{ s_i, i=i_1,...,i_m \}}}{s_{i_1}!...s_{i_m}!} z_{{\bf r_{i_1}}}^{s_{i_1}}...z_{{\bf r_{i_m}}}^{s_{i_m}},
\label{quasicumulants}
\end{equation}    
which is the most suitable for computing the unity-occupation probabilities in Eq. (\ref{rho1mTheta}), and then uses an exponential formula, or a polymer expansion, with the Bell polynomials, corresponding to that particular case of the Fa$\acute{\text{a}}$ di Bruno's formula, to compute the unity-occupation probabilities, moments, and other characteristics of the joined occupation statistics. Note that the expansion in Eq. (\ref{quasicumulants}) is different from a standard cumulant expansion 
\begin{equation}
\ln \Theta_N = \sum_{\{ {\bf r_i}, i=i_1,...,i_m \}} \sum_{s_{i_1},...,s_{i_m}} \kappa_{\{ s_i \}} \prod_{i=i_1,...,i_m} \frac{(iu_{{\bf r_i}})^{s_i}}{s_i !},  
\label{cumulants}
\end{equation}
which is more directly related to the moments of distribution, and from the generating-cumulant expansion
\begin{equation}
\ln \Theta_N =  \sum_{\{ {\bf r_i}, i=i_1,...,i_m \}} \sum_{s_{i_1},...,s_{i_m}} \tilde{\kappa}_{\{ s_i \}} \prod_{i=i_1,...,i_m} \frac{(z_{{\bf r_i}}-1)^{s_i}}{s_i !},
\label{generatingcumulants}
\end{equation}
which is the most powerful in accounting for a discreteness of stochastic variables and was introduced in \cite{PRA2010}. A set of non-negative integers $\{ s_i \}$ specifies the quasi-cumulants $\kappa'_{\{ s_i \}}$, the cumulants $\kappa_{\{ s_i \}}$, and the generating cumulants $\tilde{\kappa}_{\{ s_i \}}$ in Eqs. (\ref{quasicumulants}), (\ref{cumulants}), and (\ref{generatingcumulants}), respectively; $\kappa'_{\{ s_i =0\}}= \ln \rho_{0}^{\{N\}}\equiv \ln \langle f_0 \rangle$, $\kappa_{\{ s_i =0\}}=0$, $\tilde{\kappa}_{\{ s_i =0\}}=0$. In the case of one variable, the cumulants and generating cumulants are related by a Stirling transformation \cite{PRA2010} 
\begin{equation}
\tilde{\kappa}_m = \sum_{r=1}^{m}{S_{m}^{(r)}\kappa_r}; \quad \kappa_r = \sum_{m=1}^{r}{\sigma_{r}^{(m)}\tilde{\kappa}_m},
\label{cgc}
\end{equation}
where $S_{m}^{(r)}$ and $\sigma_{r}^{(m)}$ are the Stirling numbers of the first and second kinds, respectively. In the case of a joined distribution, a similar relation should be based on a multivariate generalization of the Stirling transformation. 

    Here we analytically compute the quasi-cumulants $\kappa'_{\{ m \}}$ and the generating cumulants $\tilde{\kappa}_{\{ m \}}$ for any combinations of unity integers $s_{\bf r}=1$ at ${\bf r} \in\{ m \} = \{ {\bf r}_{i_1},..., {\bf r}_{i_m} \}$ and zero integers $s_{\bf r'}=0$ at the rest ${\bf r'} ={\bf r}_{i_{m+1}},..., {\bf r}_{i_N}$, which correspond to the multilinear terms in the Taylor expansions (\ref{quasicumulants}), (\ref{generatingcumulants}) and are the only ones, contributing to unity-occupation probabilities in Eq. (\ref{rho1mTheta}). We use Eqs. (\ref{EqsThetaN}), (\ref{b-Theta}), written in the form
\begin{equation}
\frac{\partial \ln \Theta_N}{\partial z_{\bf r}} = -\frac{1}{z_{\bf r}} +\frac{1}{z_{\bf r}} \Big( \frac{1}{1-(1+g^{-1})z} \Big)_{1{\bf r}}^{1{\bf r}} .    
\label{derivativeTheta}
\end{equation}    
In order to evaluate it, we derived a general theorem (its proof will be published elsewhere) that yields a Taylor expansion of a diagonal element of an inverse $N\times N$-matrix $(1-Ax)^{-1}$, similar to the MacMahon master theorem in Eq. (\ref{MMM}). Namely, for $k \in \{ i_1,...,i_m \}$ we find
\begin{equation}
\frac{\partial ((1-Ax)^{-1})_k^k}{\partial x_{i_1}}\Big|_{\{ x_i =0, i=1,...,N\}} = A_k^k , \quad m=1,
\label{KMT1}
\end{equation}
\begin{equation}
\frac{\partial^m ((1-Ax)^{-1})_k^k}{\partial x_{i_1}...\partial x_{i_m}}\Big|_{\{ x_i =0\}} = \text{per}[A_{\{ m \}}-\text{diag}A_{\{ m \}}] , m \geq 2,
\label{KMT2}
\end{equation}
where a subscript ${\{ m \}}$ in $A_{\{ m \}}$ means that $(A_{\{ m \}i}^{i'})$ is a $m\times m$-matrix block with rows and columns $i,i' =i_1,...,i_m$, related to the derivatives; $\text{diag}A_{\{ m \}} = (A_i^i \delta_{i,i'})$. The derivatives in Eqs. (\ref{KMT1})-(\ref{KMT2}) are zero if $k \neq i_1,...,i_m$.

    Moreover, we generalize that theorem to the case, when the matrices $A$ and $x$ are replaced by the $N\times N$-matrices $A' =({A'}_{\bf r}^{\bf r'})$ and $z'=({z'}_{\bf r}^{\bf r'})$ (over the site-indexes ${\bf r},{\bf r'}$), whose elements themselves are the $2\times2$-matrices ${A'}_{\bf r}^{\bf r'} =(A_{j{\bf r}}^{j'{\bf r'}})$ and ${z'}_{\bf r}^{\bf r'} =(z_{\bf r} \delta_{{\bf r},{\bf r'}} \delta_{j,j'})$ over the creation/annihilation-operator indexes $j=1,2$ and $j'=1,2$. This is necessary, because a structure of the matrices $1+g^{-1}$ and $z$ in Eq. (\ref{derivativeTheta}) is exactly of that kind. For this case, we introduce a block-matrix permanent 
\begin{equation}
\text{per}_{\bf r_1}(A-\text{diag}_{\bf r}A) \equiv T_{\bf r_1} \text{per} \bar{A} = \sum_{\sigma (1,...,N)} T_{\bf r_1} \prod_{i=1,...,N} \bar{A}_{\bf r_i}^{\bf r_{\sigma(i)}}
\label{matrixper}
\end{equation}
$$= \sum_{\sigma (2,...,N)} \sum_{j_{\sigma (2)},...,j_{\sigma (N)}} \bar{A}_{j{\bf r_1}}^{j_{\sigma (2)}{\bf r_{\sigma (2)}}}\bar{A}_{j_{\sigma (2)}{\bf r_{\sigma (2)}}}^{j_{\sigma (3)}{\bf r_{\sigma (3)}}}...\bar{A}_{j_{\sigma (N)}{\bf r_{\sigma (N)}}}^{j'{\bf r_1}}$$
of the $2N\times 2N$-matrix $\bar{A}=A-\text{diag}_{\bf r} A$, that differs from $A$ by a subtracted block-diagonal part $\text{diag}_{\bf r} A = (A_{j{\bf r}}^{j'{\bf r}}\delta_{{\bf r},{\bf r'}})$. It is defined via a permanent action, similar to the one of the standard permanent in Eq. (\ref{per}), but with respect only to the indexes ${\bf r},{\bf r'}$, without involvement of the other tensor indexes $j,j'$. The block-matrix permanent is a sum of all standard products of the $N$ $2\times2$-matrices $\bar{A}_{\bf r}^{\bf r'}$. Each product begins with a $2\times2$-matrix $\bar{A}_{\bf r_1}^{\bf r_{\sigma (2)}}$ from the ${\bf r_1}$-row of matrix $\bar{A}$ and is ordered in such a way that immediately to the right of any $2\times2$-matrix $\bar{A}_{\bf r}^{\bf r_{\sigma (i)}}$ with a column-index ${\bf r_{\sigma (i)}}$ is located a $2\times2$-matrix $\bar{A}_{\bf r_{\sigma (i)}}^{\bf r'}$ with an equal row-index ${\bf r_{\sigma (i)}}$. Such index configuration reminds a standard Einstein's notation for an implicit summation, but with summation only over the index $j_{\sigma (i)}=1,2$, without summation over the index ${\bf r_{\sigma (i)}}$. This order, denoted by a symbol $T_{\bf r_1}$ in Eq. (\ref{matrixper}), is possible and unique due to the definition of the standard permanent. In a result, we find a universal formula for the multilinear quasi-cumulants
\begin{equation}
\kappa'_{\bf r_{i_1}} = \frac{1}{2} \text{Tr}_j (A_{j{\bf r_{i_1}}}^{j'{\bf r_{i_1}}}) = 1+ (g^{-1})_{1{\bf r_{i_1}}}^{1{\bf r_{i_1}}}  \quad \text{if} \ m=1 ,
\label{2x2qcumulant1}
\end{equation}
\begin{equation}
\kappa'_{\{ m \}} = \frac{1}{2} \text{Tr}_j \text{per}_{\bf r_{i_1}} (A_{\{ m \}} - \text{diag}_{\bf r} A_{\{ m \}} )  \ \text{if} \ m \geq 2,
\label{2x2qcumulants}
\end{equation}
where $A=1+g^{-1}$, a subscript ${\{ m \}}$ in $A_{\{ m \}}$ means that $(A_{\{ m \}j{\bf r}}^{j'{\bf r'}})$ is a $2m\times 2m$-matrix block, corresponding to a chosen subset of sites ${\bf r},{\bf r'} \in \{ {\bf r_{i_1}},...,{\bf r_{i_m}} \}$, $\text{diag}_{\bf r} A_{\{ m \}} =(A_{{\{ m \}}j{\bf r}}^{j'{\bf r'}} \delta_{{\bf r},{\bf r'}})$ is a diagonal part of $A_{\{ m \}}$ over indexes ${\bf r},{\bf r'}$, and $\text{Tr}_j$ denotes a trace of a subsequent matrix over the creation-annihilation indexes $j=1,2$ and $j'=1,2$.

    The universal analytical result for the multilinear quasi-cumulants in Eqs. (\ref{2x2qcumulant1}), (\ref{2x2qcumulants}) allows one to compute the unity-occupation probabilities in Eq. (\ref{rho1mTheta}) by means of the exponential formula. In particular, these joined unconstrained distributions for unity occupations of spin bosons at one, two, or three sites are
\begin{equation}
 \rho_{1{\bf r_1}} = \kappa'_{\bf r_1} /\sqrt{\text{det}g} = [1+ (g^{-1})_{1{\bf r_1}}^{1{\bf r_1}}]/\sqrt{\text{det}g} ,
\label{rho1-1}
\end{equation}    
\begin{equation}    
 \rho_{1{\bf r_1}{\bf r_2}} = (\kappa'_{\bf r_1}\kappa'_{\bf r_2}+\kappa'_{{\bf r_1}{\bf r_2}})/\sqrt{\text{det}g} ,
\label{rho1-2}
\end{equation}  
\begin{equation}
 \rho_{1{\bf r_1}{\bf r_2}{\bf r_3}} 
\label{rho1-3}
\end{equation}
$$= \frac{\kappa'_{\bf r_1}\kappa'_{\bf r_2}\kappa'_{\bf r_3} +\kappa'_{\bf r_1}\kappa'_{{\bf r_2}{\bf r_3}} +\kappa'_{\bf r_2}\kappa'_{{\bf r_1}{\bf r_3}} + \kappa'_{\bf r_3}\kappa'_{{\bf r_1}{\bf r_2}} +\kappa'_{{\bf r_1}{\bf r_2}{\bf r_3}}}{\sqrt{\text{det}g}}.$$

    The presented analysis of the multilinear quasi-cumulants $\kappa'_{\{ m \}}$ can be fully transfered to a similar analysis of the multilinear generating cumulants $\tilde{\kappa}_{\{ m \}}$, if one uses the expansion in Eq. (\ref{generatingcumulants}) over $(z_{{\bf r_i}}-1)$ instead of the  expansion in Eq. (\ref{quasicumulants}) over $z_{{\bf r_i}}$ and the following representation of the characteristic function
\begin{equation}
\Theta_N (\{ u_{\bf r} \}) = \frac{1}{\sqrt{\text{det}[1-\tilde{A}(z-1)]}}, \quad  \tilde{A}=-(1+g) ,
\label{ThetaNgenerating}
\end{equation}
instead of the one in Eq. (\ref{ThetaNresult}). In this way, an exact result for the multilinear generating cumulants immediately follows from Eqs. (\ref{2x2qcumulant1}), (\ref{2x2qcumulants}) by means of replacement of the matrix $A$ by the matrix $\tilde{A}=-(1+g)$, that is
\begin{equation}
\tilde{\kappa}_{\bf r_{i_1}} = \frac{1}{2} \text{Tr}_j (\tilde{A}_{j{\bf r_{i_1}}}^{j'{\bf r_{i_1}}}) = -1- g_{1{\bf r_{i_1}}}^{1{\bf r_{i_1}}}  \quad \text{if} \ m=1 ,
\label{2x2gcumulant1}
\end{equation}
\begin{equation}
\tilde{\kappa}_{\{ m \}} = \frac{1}{2} \text{Tr}_j \text{per}_{\bf r_{i_1}} (\tilde{A}_{\{ m \}} - \text{diag}_{\bf r} \tilde{A}_{\{ m \}} )  \quad \text{if} \ m \geq 2.
\label{2x2gcumulants}
\end{equation}

     For the joined unconstrained, non-cutoff distribution of the spin-bosons' occupations at only a subset of lattice sites $\{ M \}=\{{\bf  r_k}, k=1,...,M\}, M\leq N$, defined in Eq. (\ref{rhoM}), we follow an analogy with derivation of Eq. (\ref{ThetaNresult}) and repeat the steps in Eqs. (\ref{EqsThetaN})-(\ref{ThetaNresult}), restricting, in accord with the argument in Eq. (\ref{2mx2m}), the general solution for the partial operator contraction in Eq. (\ref{b-solution}) by the only relevant quasi-diagonal $2M\times 2M$-block for the subset of sites $\{ M \}=\{{\bf  r_k}, k=1,...,M\}$, the corresponding quasi-diagonal $2M\times 2M$-block $g_{\{M\}}$ of the correlation $2N\times 2N$-matrix $g$, and the $2M\times 2M$-block 
\begin{equation}
Z^{\{M\}}_{\Theta}= \text{diag}\{(Z_{\Theta})_I^I ;{\bf r}={\bf r_1},...,{\bf r_M}; j=1,2\}\equiv \frac{z_{\{M\}}}{z_{\{M\}}-1}
\label{ZThetaM}
\end{equation}
of the diagonal matrix $Z_{\Theta}$ in Eq. (\ref{ZTheta}). The result
$$\Theta^{\{M\}}_N (\{ u_{\bf r} \}) = \frac{1}{\sqrt{\text{det}(g_{\{M\}}+Z^{\{M\}}_{\Theta})}} \prod_{{\bf r}={\bf r_1}}^{\bf  r_M} \frac{1}{1-z_{\bf r}}$$ 
\begin{equation}
= \frac{1}{\sqrt{\text{det}g_{\{M\}}}} \frac{1}{\sqrt{ \text{det}(1-(1+g_{\{M\}}^{-1})z_{\{M\}})}}
\label{ThetaMresult}
\end{equation}
is similar to Eq. (\ref{ThetaNresult}). Obviously, its differentiation, 
\begin{equation}
\rho^{\{M\}}_{1\{m\}}= \frac{\partial^m \Theta^{\{M\}}_N}{\partial z_{\bf r_1}... \partial z_{\bf r_m}} \Big|_{\{z_{\bf r}=0\}} ,
\label{rho1mMTheta}
\end{equation}     
yields the corresponding, similar to Eq. (\ref{rho1mTheta}), unconstrained probabilities for spin bosons to have unity occupations at subset of lattice sites $\{ m \} =\{{\bf  r_i}, i=1,...,$ $m\}, m\leq M \leq N$, zero occupations at subset of lattice sites $\{ M \} \setminus \{ m \} =\{{\bf  r_k}, k=m+1,...,M \}$, and arbitrary occupations, $n_{\bf r'}=0,1,2,...,\infty$, at the rest $N-M$ sites. 

    The analysis and corresponding results for these probabilities $\rho^{\{M\}}_{1\{m\}}$ and related quasi-cumulants ${\kappa'}_{\{ m \}}^{\{ M \}}$ and generating cumulants $\tilde{\kappa}_{\{ m \}}^{\{ M \}}$ are exactly the same as the ones, presented above in Eqs. (\ref{rho1mTheta})-(\ref{2x2gcumulants}), if one replaces the $2N\times 2N$ equal-time anti-normally ordered correlation matrix $g$ by its $2M\times 2M$-version $g_{\{ M \}}$, restricted to the subset of sites $\{ M \}$. We do not repeat them here. 

     The derived characteristic functions in Eqs. (\ref{ThetaNresult}) and (\ref{ThetaMresult}) immediately yield the probability for all $N$ or for a subset $\{ M \}=\{{\bf  r_k}, k=1,...,M\}, M\leq N$, of spin bosons to have zero occupations as the value $\Theta^{\{M\}}_N |_{\{ z_{\bf r}=0\}}$:
\begin{equation}
\rho_{0}^{\{N\}}\equiv \langle f_0 \rangle=\frac{1}{\sqrt{\text{det}g}}, \quad \rho_{0}^{\{M\}}\equiv \langle f_0^{\{M\}} \rangle = \frac{1}{\sqrt{\text{det} g_{\{ M\} }} }.  
\label{rho0N}
\end{equation} 

     Finally, we present the explicit formulas for the characteristic functions of the joined non-cutoff probability distributions in the most interesting cases of the single-site ($M=1$) and two-sites ($M=2$) subsets of spin bosons:
\begin{equation}     
\Theta^{\{1\}}_N (u_{\bf r}) = \frac{\rho_0}{\sqrt{\text{det}(1-(1+S^{-1})z_{\{1\}})}} ,   
\label{Theta1}
\end{equation}
\begin{equation}
\Theta^{\{2\}}_N (u_{\bf r_1},u_{\bf r_2}) =\frac{\rho_{0,0}}{\sqrt{\text{det}(1-(1+q^{-1})z_{\{2\}})}} . 
\label{Theta2}
\end{equation}
For the corresponding probabilities of zero occupations we have the following compact results in terms of the determinants of the auto- and cross-correlation matrices, defined in Eq. (\ref{q4x4}), 
\begin{equation}
\rho_0\equiv \rho_{0}^{\{1\}} = 1/\sqrt{\text{det}S} , \quad
\rho_{0,0}\equiv \rho_{0}^{\{2\}} = 1/\sqrt{\text{det}q} . 
\label{rho0M1}
\end{equation}
For the single-site probability one has $M=1$ and $2\times 2$-matrix $g_{\{1\}}=S$, Eq. (\ref{q4x4}), so that Eqs. (\ref{rho1mMTheta}), (\ref{Theta1}) yield
\begin{equation}
\rho_1\equiv \rho_{1\{1\}}^{\{1\}} = \rho_0 \Big[1+ (S^{-1})_{1{\bf r}}^{1{\bf r}}\Big] \equiv \rho_0 \Big[1+ \frac{g_{1{\bf r}}^{1{\bf r}}}{\text{det}S}\Big] .
\label{rho1/rho0}
\end{equation}
For the two-sites probabilities one has $M=2$ and $4\times 4$-matrix $g_{\{2\}}=q$ (where hereinafter a matrix $q_{j{\bf R}}^{j'{\bf R'}}$ is defined similar to Eq. (\ref{q4x4}) with the ${\bf R}$ and ${\bf R'}$ running over arbitrary two sites $\{{\bf  r_1}, {\bf  r_2} \}$, not necessarily neighboring sites), and Eqs. (\ref{rho1mMTheta}), (\ref{Theta2}), in accord with Eq. (\ref{rho1-2}), yield
\begin{equation}
\rho_{1,0}\equiv \rho_{1\{1\}}^{\{2\}} = \rho_{0,0} \Big[1+ (q^{-1})_{1{\bf r_1}}^{1{\bf r_1}}\Big] \equiv \rho_{0,0} \Big[1+ \frac{\text{det}[q]_{1{\bf r_1}}^{1{\bf r_1}}}{\text{det}q}\Big],
\label{rho10/rho00}
\end{equation}
\begin{equation}
\rho_{0,1}\equiv \rho_{1\{1\}}^{\{2\}} = \rho_{0,0} \Big[1+ (q^{-1})_{1{\bf r_2}}^{1{\bf r_2}}\Big] \equiv \rho_{0,0} \Big[1+ \frac{\text{det}[q]_{1{\bf r_2}}^{1{\bf r_2}}}{\text{det}q}\Big],
\label{rho01/rho00}
\end{equation}

$$\rho_{1,1}\equiv \rho_{1\{2\}}^{\{2\}} = \rho_{0,0} \Big\{ \Big[1+ (q^{-1})_{1{\bf r_1}}^{1{\bf r_1}}\Big] \Big[1+ (q^{-1})_{1{\bf r_2}}^{1{\bf r_2}}\Big] $$
\begin{equation}
+ (q^{-1})_{1{\bf r_1}}^{1{\bf r_2}}(q^{-1})_{1{\bf r_2}}^{1{\bf r_1}} + (q^{-1})_{1{\bf r_1}}^{2{\bf r_2}}(q^{-1})_{2{\bf r_2}}^{1{\bf r_1}} \Big\} ,
\label{rho11/rho00}
\end{equation}
where $[q]_{I}^{I'}$ stands for a $II'$-submatrix, i.e., matrix $q$ with $I$-th row and $I'$-th column deleted. Together with the probabilities $\rho_0\equiv \rho_{0}^{\{1\}}$ and $\rho_{0,0}\equiv \rho_{0}^{\{2\}}$ of zero relevant occupations, given in Eq. (\ref{rho0M1}), these results yield the probabilities, requested by self-energy Eqs. (\ref{pSelfEnergyIsing}), (\ref{newSigma(l)}). Note that the formulas for the characteristic functions (108) and (130) in \cite{PhysicaScripta2015} as well as their copies (18) and (24) in \cite{PLA2015Ising} contained misprints and, together with the following from them formulas for corresponding two-sites probabilities $\rho_{n_1,n_2}$, should be replaced by the present correct Eqs. (\ref{ThetaMresult}), (\ref{Theta2})-(\ref{rho11/rho00}) for probabilities $\rho_{n_1,n_2}$. Those formulas served only as illustrations and were not important for the main subject of papers \cite{PhysicaScripta2015,PLA2015Ising} - a general method of solution for 3D Ising model. 

     In fact, the full non-cutoff distributions of spin-boson occupations, neither single-site nor joined, are not explicitly required for the solution of 3D Ising model. Nevertheless, in order to illustrate the technique of recurrence equations (\ref{spin2-contraction}) for partial operator contractions, we outline a full single-site non-cutoff distribution of spin-boson occupations. It can be expressed via a Jacobi polynomial $P_n^{(\alpha,\beta)}$ or a hypergeometric function, i.e., a generalized hypergeometric series $F(a,b;c;z)=\sum_{k=0}^{\infty} \frac{(a)_k(b)_kz^k}{(c)_k k!}$, as
\begin{equation}
\rho_n =\frac{\rho_0}{\zeta_1^n}P_n^{(0,-n-\frac{1}{2})} (\frac{2\zeta_1}{\zeta_2}-1) =\frac{\rho_0}{\zeta_2^n}F(-n,\frac{1}{2};1;1-\frac{\zeta_2}{\zeta_1}),
\label{rhoJacobi}
\end{equation}
where $\zeta_{1,2} =\frac{g_{1{\bf r}}^{1{\bf r}}\mp |g_{1{\bf r}}^{2{\bf r}}|}{1+g_{1{\bf r}}^{1{\bf r}} \mp |g_{1{\bf r}}^{2{\bf r}}|}$. We apply Eq. (\ref{spin2-contraction}) to the two-operator contraction $\tilde{b}_{1{\bf r}}^{1{\bf r}} [\delta_{\tilde{n}_{\bf r}-1, n}]$, which is related to the single-site non-cutoff distribution by an equation
\begin{equation}
\rho_n \equiv \langle \delta_{\tilde{n}_{\bf r},n} \rangle = \frac{\langle \tilde{b}_{1{\bf r}}^{1{\bf r}}[\delta_{\tilde{n}_{\bf r}-1,n}] \rangle}{n+1}, \ n=0,1,2, \dots .
\label{rho-b}
\end{equation}
The result is the following recurrence equation for $\rho_n$:
\begin{equation}
\rho_{n+2} -\frac{2n+3}{n+2}[1+(S^{-1})_{1{\bf r}}^{1{\bf r}}]\rho_{n+1} +\frac{n+1}{n+2}\text{det}(1+\frac{1}{S})\rho_n =0.
\label{rho-recurrence}
\end{equation} 
It is of a Jacobi polynomial's type, and we immediately get the result in Eq. (\ref{rhoJacobi}).

    Thus, we obtain an exact analytical solution in Eqs. (\ref{ThetaNresult}) and (\ref{ThetaMresult}) for the joined unconstrained statistics of spin-bosons' occupations. It has a universal structure in terms of the equal-time anti-normally ordered correlation matrix $g$ and related determinants, that allows one to use the powerful methods of matrix algebra for its explicit analysis and calculation. In particular, we employ a well-known MacMahon master theorem \cite{Percus1971} on the inverse determinant in Eq. (\ref{MMM}), a similar theorem on the inverse matrix in Eqs. (\ref{KMT1})-(\ref{KMT2}), the permanents in Eq. (\ref{per}), and the block-matrix permanents in Eq. (\ref{matrixper}) to compute analytically the quasi-cumulants in Eqs. (\ref{2x2qcumulant1})-(\ref{2x2qcumulants}), the generating cumulants in Eqs. (\ref{2x2gcumulant1})-(\ref{2x2gcumulants}), and the joined probabilities of configurations with no more than one quantum of excitation in each spin boson of a given subset of lattice sites in Eqs. (\ref{rho1mTheta}), (\ref{rho1per}), (\ref{rho1-1})-(\ref{rho1-3}), (\ref{rho1mMTheta})-(\ref{rho11/rho00}). The exact solutions for those unity-occupation probabilities are precisely the results, which are required for finding the exact solution of 3D Ising model.

\section{The true order parameter, correlation functions, and joined statistics of spin-bosons' occupations}

    The true, constrained thermal average, defined as 
\begin{equation}
\langle \dots \hat{\theta} \rangle /P_s \equiv  \text{Tr}\{\dots \hat{\theta} e^{-\frac{H}{T}}\}/(P_s\text{Tr}\{e^{-\frac{H}{T}}\}), \ P_s= \langle \hat{\theta} \rangle ,
\label{trueaverage}
\end{equation}
differs from the non-cutoff, auxiliary average in Eq. (\ref{average}) by a presence of the product of $N$ cutoff factors $\hat{\theta} =  \prod_{\bf r} \theta(2s-\hat{n}_{\bf r})$ and by the normalization factor $P_s= \langle \hat{\theta} \rangle $. That true average should be used for calculation of all observable quantities as well as the true, constrained Green's functions, defined in Eq. (\ref{spinGreen'}). In particular, the true average of the $z$-component of a spin in Eq. (\ref{spincomponents}), 
\begin{equation}
\bar{S}^{'z}_{\bf r} = s- \langle \hat{n}_{\bf r}\hat{\theta} \rangle/P_s ,
\label{orderIsing}
\end{equation}
represents a magnetization, that is an order parameter at a lattice site ${\bf r}$ within the Ising model. For the spin value $s=1/2$, it is reduced to a quantity
\begin{equation}
\bar{S}^{'z}_{\bf r} = 1/2 - \rho'_{n_{\bf r}=1} , \qquad \rho'_{n_{\bf r}=n} =\langle \delta_{\hat{n}_{\bf r},n}\hat{\theta} \rangle /P_s , 
\label{spinIsing}
\end{equation}
determined by a $\hat{\theta}$-cutoff, true probability $\rho'_{n_{\bf r}=1}$ of a spin boson at site ${\bf r}$ to have one quantum of excitation. 

    The true, constrained correlation functions of spin bosons in a lattice ${g'}_{I_1}^{I_2}$ can be found as the Green's functions $G^{'J_2}_{J_1}$ in Eq. (\ref{spinG'}) in the equal-time limit $\tau_1 \to \tau_2 -(-1)^{j_2}\times 0$. The partial two-operator contraction, required by Eq. (\ref{spinG'}), is given by the exact solution in Eq. (\ref{b-solution}) for $f=\hat{\theta}$. The only nonzero terms in the latter solution come from the lowest terms of its expansion over matrix $Z$, 
 $\tilde{b}[\hat{\theta}]= -Z+Zg^{-1}Z+...$, and only from the first term in the  geometrical-progression expansion of $Z= -d-...$ in Eq. (\ref{inverse(1-D)}) over the raising-operator matrix $d=D^{-1}$. In a result, the exact solution for the required two-operator contraction is reduced to a simple form
\begin{equation}   
\tilde{b}^{I_2}_{I_1} = [(g^{-1})_{I_1}^{I_2} (1-\delta_{{\bf r_1},{\bf r_2}}) \theta(-\tilde{n}_{{\bf r_1}}) \theta(-\tilde{n}_{{\bf r_2}})+
\delta_{I_1,I_2}\theta(-\tilde{n}_{{\bf r_1}}) ]\hat{\theta} 
\label{b(theta)Ising}
\end{equation}
that yields an exact analytical solution for the true, constrained correlation function of spin bosons in a lattice 
\begin{equation}    
{g'}_{I_1}^{I_2} = -(g^{-1})_{I_1}^{I_2} (1-\delta_{{\bf r_1},{\bf r_2}}) \rho'_{n_{\bf r_1}=0,n_{\bf r_2}=0} -\rho'_{n_{\bf r_1}=0} \delta_{I_1,I_2}.
\label{g'Ising}
\end{equation} 
The matrix $g^{-1}$, which is inverse to the matrix $(g_{I_1}^{I_2})$ of unconstrained correlations, can be calculated by a technique of Toeplitz matrices, known from theory of 2D Ising model \cite{Kadanoff,Montroll1963,Wu1976}. Eq. (\ref{g'Ising}) involves a true probability
\begin{equation} 
\rho'_{n_{\bf r_1}=n_1,n_{\bf r_2}=n_2} =\langle \delta_{\hat{n}_{\bf r_1},n_1}\delta_{\hat{n}_{\bf r_2},n_2}\hat{\theta} \rangle /{P_s}
\label{true2sidesprob}
\end{equation}
for two spin-bosons at sites ${\bf r_1}$ and ${\bf r_2}$ to have $n_{\bf r_1}=n_1$ and $n_{\bf r_2}=n_2$ quanta of excitation, respectively. Namely, it involves a true probability for two spin-bosons to have zero quanta of excitation $n_1=n_2=0$ simultaneously. 

    Next, we present the exact analytical formulas for such true joined distributions $\rho'_{n_{\bf r}}$, $\rho'_{n_{\bf r_1},n_{\bf r_2}}$, $\rho'_{\{ n_{\bf r}\}}$ of physically allowable spin-boson occupations ${n_{\bf r}}=0$ or ${n_{\bf r}}=1$. Those distributions are simply the $\hat{\theta}$-cutoff versions of the calculated in Sec. IV unconstrained distributions $\rho_{n_{\bf r}}$, $\rho_{n_{\bf r_1},n_{\bf r_2}}$, $\rho_{\{ n_{\bf r}\}}$, restricted to the unity-occupation ones $\rho_{1\{m\}}$ in Eq. (\ref{rho1mTheta}). Note that the unconstrained joined distributions of spin-bosons' occupations already contain all effects of the constraints and spin interaction, except the $\hat{\theta}$-cutoff only, since they were calculated for the exact, constrained and $\hat{\theta}$-cutoff, Hamiltonian (\ref{HIsing}). 
    
    We start the analysis of the true joined distribution of the occupations $\{n_{\bf r}=0 \ \text{or} \ 1\}$ for all $N$ spin bosons, 
\begin{equation}
\rho'_{\{ n_{\bf r}\}} \equiv \frac{1}{P_s} \langle \prod_{\bf r}\delta_{\hat{n}_{\bf r},n_{\bf r}}\hat{\theta} \rangle , \quad P_s= \langle \hat{\theta} \rangle ,
\label{rho'N}
\end{equation}
with an evaluation of the normalization factor $P_s$. It is equal to the sum of the probabilities $\rho_{1\{m\}}$ in Eq. (\ref{rho1mTheta}) over all occupation configurations $\{ {n_{\bf r}}=0 \ \text{or} \ 1; \ {\bf r} ={\bf r_1},...,{\bf r_N} \}$, which can be written as 
$$P_s = \frac{\partial^N}{\partial z_{\bf r_1}... \partial z_{\bf r_N}} \Big[ \Theta_N \prod_{{\bf r}={\bf r_1},...,{\bf r_N}} (1+z_{\bf r}) \Big] \Big|_{\{z_{\bf r}=0\}}$$
\begin{equation}
= \frac{\partial^N \Theta'}{\partial z_{\bf r_1}... \partial z_{\bf r_N}} \Big|_{\{z_{\bf r}=0\}}, \ \Theta' = \frac{1}{\sqrt{\text{det}g}} \frac{1}{\sqrt{\text{det}[1-(2+g^{-1})z]}} .
\label{partialPs}
\end{equation}
The second equality in the equation for $P_s$ is due to the fact, that the terms with the square, $z_{\bf r}^2$, and higher powers of any variable $z_{\bf r}$ do not contribute to the considered derivative at the zero point $\{ z_{\bf r}=0 \}$. Note that the newly introduced function $\Theta'$ differs from the characteristic function $\Theta_N$ in Eq. (\ref{ThetaNresult}) only by a substitution of the matrix $A=1+g^{-1}$ with the matrix 
\begin{equation}
A'' = 2 +g^{-1} .
\label{A'}
\end{equation}
Thus, an evaluation of the normalization factor $P_s$ can be done similar to the evaluation of probability $\rho_{1\{m\}}$ at $m =N$, described in Sec. IV. In particular, in the case of vanishing anomalous correlations $g_{1{\bf r}}^{2{\bf r'}}=0$ and non-zero normal correlations $g_{1{\bf r}}^{1{\bf r'}} =g_{2{\bf r}}^{2{\bf r'}}$, as in Eq. (\ref{rho1per}), we find 
\begin{equation}
P_s= \frac{\text{per} (2+g_1^{-1})}{\text{det}g_1}, \quad g_1 \equiv g_{1\{ N \}} \equiv [g_{1{\bf r}}^{1{\bf r'}}] .
\label{perP_s}
\end{equation}

     A result for the single-site zero occupation probability 
\begin{equation}
\rho'_{ n_{\bf r_1}=0}= \frac{1}{P_s}\frac{\partial^{N-1} \Theta'}{\partial z_{\bf r_2}... \partial z_{\bf r_N}} \Big|_{\{z_{\bf r}=0\}}
\label{rho'1-0}
\end{equation}
differs from $P_s$ only by an absence of one partial derivative $\partial /\partial z_{\bf r_1}$ and by a normalization factor. In particular, in the case of vanishing anomalous correlations $g_{1{\bf r}}^{2{\bf r'}}=0$ and non-zero normal correlations $g_{1{\bf r}}^{1{\bf r'}} =g_{2{\bf r}}^{2{\bf r'}}$, we have
\begin{equation}
\rho'_{ n_{\bf r_1}=0} =\frac{\text{per} (2+g_1^{-1})_{\{N-1 \}}}{\text{per} (2+g_1^{-1})}; 
\label{rho'1-0per}
\end{equation}
$$(2+g_1^{-1})_{\{N-1 \}}\equiv ((2+g_1^{-1})_{\bf r}^{\bf r'}), 
\quad {\bf r}, {\bf r'} \neq {\bf r_1}.$$
The true single-site unity occupation probability is equal
\begin{equation}
\rho'_{ n_{\bf r_1}=1}= 1 - \rho'_{ n_{\bf r_1}=0} .
\label{rho'1-1}
\end{equation}

    The true two-sites zero occupation probability 
\begin{equation}
\rho'_{ n_{\bf r_1}=0, n_{\bf r_2}=0}= \frac{1}{P_s}\frac{\partial^{N-2} \Theta'}{\partial z_{\bf r_3}... \partial z_{\bf r_N}} \Big|_{\{z_{\bf r}=0\}}
\label{rho'2-0}
\end{equation}
differs from the single-site one in Eq. (\ref{rho'1-0}) only by an absence of one more partial derivative $\partial /\partial z_{\bf r_2}$. So, in the case of vanishing anomalous correlations and non-zero normal correlations $g_{1{\bf r}}^{1{\bf r'}} =g_{2{\bf r}}^{2{\bf r'}}$, as in Eq. (\ref{rho1per}), one has
\begin{equation}
\rho'_{ n_{\bf r_1}=0, n_{\bf r_2}=0}=\frac{\text{per} (2+g_1^{-1})_{\{N-2 \}}}{\text{per} (2+g_1^{-1})}; 
\label{rho'2-0per}
\end{equation}
$$(2+g_1^{-1})_{\{N-2 \}}\equiv ((2+g_1^{-1})_{\bf r}^{\bf r'}), 
\quad {\bf r}, {\bf r'} \in \{ {\bf r_3},..., {\bf r_N} \}.$$
The true two-sites probabilities for other occupation combinations can be computed from the probabilities, presented above, as follows 
$$\rho'_{ n_{\bf r_1}=0, n_{\bf r_2}=1}= \rho'_{ n_{\bf r_1}=0} - \rho'_{ n_{\bf r_1}=0,  n_{\bf r_2}=0} ,$$
\begin{equation}
\rho'_{ n_{\bf r_1}=1, n_{\bf r_2}=1}= \rho'_{ n_{\bf r_1}=1} - \rho'_{ n_{\bf r_1}=1, n_{\bf r_2}=0}.
\label{rho'2-1}
\end{equation}
These equations stem from a fact that the true single-site occupation distribution is equal to the true two-sites occupation distribution, averaged over physically allowable occupations $n_{\bf r_2}=0,1$ of a spin boson at the second site:
\begin{equation}
\rho'_{ n_{\bf r_1}}=\rho'_{n_{\bf r_1},n_{\bf r_2}=0}+\rho'_{n_{\bf r_1},n_{\bf r_2}=1}.
\label{rho'1=averagerho'2}
\end{equation}

    The true three-sites and other multiple-sides occupation distributions are not required for calculation of the true order parameter and correlation functions, but are necessary for the analysis of the true multiple-sides correlations and statistics. Those m-sides occupation distributions can be computed in a similar way by induction:
\begin{equation}
\rho'_{ n_{\bf r_1}=0,..., n_{\bf r_m}=0}= \frac{1}{P_s}\frac{\partial^{N-m} \Theta'}{\partial z_{\bf r_{m+1}}... \partial z_{\bf r_N}} \Big|_{\{z_{\bf r}=0\}} , \ m \leq N,
\label{rho'm-0}
\end{equation}
\begin{equation}
\rho'_{ n_{\bf r_1},...,  n_{\bf r_{m-1}}, n_{\bf r_m}=1}= \rho'_{ n_{\bf r_1},...,  n_{\bf r_{m-1}}} - \rho'_{n_{\bf r_1},...,  n_{\bf r_{m-1}}, n_{\bf r_m}=0} .
\label{rho'm-1}
\end{equation}
In the case of vanishing anomalous correlations $g_{1{\bf r}}^{2{\bf r'}}=0$ and non-zero normal correlations $g_{1{\bf r}}^{1{\bf r'}} =g_{2{\bf r}}^{2{\bf r'}}$, one has
\begin{equation}
\rho'_{ n_{\bf r_1}=0,..., n_{\bf r_m}=0}=\frac{\text{per} (2+g_1^{-1})_{\{N-m \}}}{\text{per} (2+g_1^{-1})}; 
\label{rho'm-0per}
\end{equation}
$$(2+g_1^{-1})_{\{N-m \}}\equiv ((2+g_1^{-1})_{\bf r}^{\bf r'}), 
\ {\bf r}, {\bf r'} \in \{ {\bf r_{m+1}},..., {\bf r_N} \}.$$

    We stress, that the true joined distributions of spin-bosons' occupations, even for a subset of lattice sites $\{ M \}=\{{\bf  r_k}, k=1,...,M\}, M\leq N$, always are determined by the full $2N\times 2N$-matrix $g^{-1}$, which is inverse to the $2N\times 2N$ equal-time anti-normally ordered correlation matrix $g$. This is in contrast with the unconstrained joined distributions in Eqs. (\ref{ThetaMresult})-(\ref{rho11/rho00}), which are determined only by the corresponding quasi-diagonal $2M\times 2M$-block $g_{\{M\}}$ of the full $2N\times 2N$-matrix $g$. 
    
    A detailed analysis of the obtained true spin-bosons' occupation distributions as well as the true order parameter and correlation functions in Eqs. (\ref{spinIsing}) and (\ref{g'Ising}) will be given elsewhere, since they are not required for the self-consistency equation in Sec. VI.

\section{Exact closed self-consistency equation for the nearest-neighbors' normal and anomalous correlations}

    Now we can make a final, crucial step in the exact solution of the 3D Ising model - find an exact closed self-consistency equation for the nearest-neighbors', basis normal and anomalous auto- and cross-correlations $g_{1{\bf r_0}}^{1{\bf r_l}}=g_{2{\bf r_0}}^{2{\bf r_l}*}, \ g_{1{\bf r_0}}^{2{\bf r_l}}=g_{2{\bf r_0}}^{1{\bf r_l}*}, l=0,1,...,p$, in Eq. (\ref{q4x4}). Indeed, the total irreducible self-energy in Eqs. (\ref{pSelfEnergyIsing}), (\ref{newSigma(l)}) and the spin-bosons' unconstrained occupation probabilities, entering formulas for self-energy, in Eqs. (\ref{rho1/rho0})-(\ref{rho11/rho00}), (\ref{rho0M1}) are known exactly via the $(1+p)$ basis correlation $2\times2$-matrices $g(l), l=0,1,...,p$, Eq. (\ref{q4x4}): the matrix $S \equiv g(0) = (g_{j{\bf r_0}}^{j'{\bf r_0}})$ of auto-correlations for a spin boson at site ${\bf r_0}$ and the coordination number $p$ matrices $C(l) \equiv g(l\ne 0) = (g_{j{\bf r_0}}^{j'{\bf r_l}})$ of cross-correlations of a spin boson at site ${\bf r_0}$ with the nearest-neighbors at sites ${\bf r_l}={\bf r_0}+{\bf l}$. In fact, due to the complex-conjugation relations
\begin{equation}
g_{1{\bf r_0}}^{1{\bf r_0}}=g_{2{\bf r_0}}^{2{\bf r_0}}, \ g_{1{\bf r_0}}^{2{\bf r_0}}=g_{2{\bf r_0}}^{1{\bf r_0}*}, \ g_{1{\bf r_0}}^{1{\bf r_l}}=g_{2{\bf r_0}}^{2{\bf r_l}*}, \ g_{1{\bf r_0}}^{2{\bf r_l}}=g_{2{\bf r_0}}^{1{\bf r_l}*},
\label{cc-relations}
\end{equation}
there are only two independent, normal $g_{1{\bf r_0}}^{1{\bf r_l}}$ and anomalous $g_{1{\bf r_0}}^{2{\bf r_l}}$, correlation parameters per each basis correlation $2\times2$-matrix, that is, only $2(1+p)$ numbers, which determine all details of critical phenomena. 

    Thus, we can find the self-consistency equation for those $2(1+p)$ basis auto- and cross-correlations in two steps. First, we solve the Dyson-type Eq. (\ref{spinGEq}) for the unconstrained Green's functions in terms of those basis correlations. Second, we close the loop by expressing the basis correlations themselves via those Green's functions. 

    For the considered stationary homogeneous phases, the Green's functions, the equal-time correlation functions, and the self-energy depend only on the differences of their arguments $\tau= \tau_1-\tau_2$ and ${\bf r}= {\bf r_2}-{\bf r_1}$, that is
\begin{equation}
G_{J_1}^{J_2}=G_{j_1j_2}(\tau, {\bf r}), \ g_{j_1{\bf r_1}}^{j_2{\bf r_2}}=g_{j_1j_2}({\bf r}), \ \Sigma_{J_1}^{J_2}=\delta(\tau)\Sigma_{j_1j_2}({\bf r}). 
\label{r-r}
\end{equation}
Hence, it is straightforward to solve the Dyson-type Eq. (\ref{spinGEq}) by means of the Fourier transformation over imaginary time $\tau \in [-\frac{1}{T},\frac{1}{T}]$ and the discrete Fourier transformation over space. The latter has a following form
\begin{equation}
g({\bf k})=\sum_{\bf r}g({\bf r})e^{-i{\bf k}{\bf r}}, \ g({\bf r})= \Big( \frac{a}{L} \Big)^d \sum_{\bf k}g({\bf k})e^{i{\bf k}{\bf r}}, 
\label{DFT}
\end{equation}
where the sums run over all lattice sites  ${\bf r}$ with a period $a$ and discrete wave vectors ${\bf k}=\{ k_i; i=1,\dots,d\},k_i=\frac{2\pi}{L}q$ with an integer $q$; $k_i \in [-\frac{\pi}{a},\frac{\pi}{a}]$. We discern the Fourier transform and its inverse by the arguments ${\bf k}$ and ${\bf r}$. A result for the normal and anomalous Green's functions is
\begin{equation}
G_{11}(\tau,{\bf k})= \sum_{j=1}^{2} (-1)^j \frac{[i\omega^{(j)}+\varepsilon+\Sigma_{22}({\bf k})]e^{i\omega^{(j)}(\frac{\text{sign}(\tau)}{2T}-\tau)}}{2(\omega^{(2)}-\omega^{(1)})\sin [\omega^{(j)}/(2T)]} ,
\label{G11}
\end{equation}
\begin{equation}
G_{12}(\tau,{\bf k})= \sum_{j=1}^{2} \frac{(-1)^j \Sigma_{12}({\bf k})e^{i\omega^{(j)}(\frac{\text{sign}(\tau)}{2T}-\tau)}}{2(\omega^{(1)}-\omega^{(2)})\sin [\omega^{(j)}/(2T)]} ,
\label{G12}
\end{equation}
where the two quasiparticle eigen-energies 
\begin{equation}
i\omega^{(1,2)}= \frac{\Sigma_{11}-\Sigma_{22}}{2} \pm \Big[ \Big( \varepsilon +\frac{\Sigma_{11}+\Sigma_{22}}{2} \Big)^2 -\Sigma_{12}\Sigma_{21} \Big]^{\frac{1}{2}} 
\label{omega12}
\end{equation}
depend on the wave vector ${\bf k}$ via the self-energies 
\begin{equation}
\Sigma_{j_0j}({\bf k})= \sum_{l=0}^p \Sigma_{j_0}^{j}(l) e^{-i{\bf k}({\bf r_l}-{\bf r_0})} , \quad {\bf r_l}={\bf r_0}+{\bf l}. 
\label{Sigma(k)}
\end{equation}
The latter Fourier transform of the self-energy consists of only $1+p$ terms within a neighborhood of the nearest sites, for which there are nonzero couplings $J_{{\bf r_0},{\bf r_l}} \neq 0$. This is a consequence of the fact, that, according to Eq. (\ref{pSelfEnergyIsing}), the self-energy matrix is a diagonal, $2(p+1)$-banded matrix. The $2\times2$-matrix blocks $\Sigma(l)$ are given explicitly in Eqs. (\ref{p+1SelfEnergyIsing})-(\ref{Sigma(l)}), derived from Eqs. (\ref{pSelfEnergyIsing}), (\ref{newSigma(l)}).

    The spatial Fourier transforms of the normal and anomalous equal-time correlation functions follow from Eqs. (\ref{G11}) and (\ref{G12}) in the limit $\tau \to +0$:
\begin{equation} 
g_{11}({\bf k})= \sum_{j=1}^{2} \frac{(-1)^j [i\omega^{(j)}+\varepsilon+\Sigma_{22}({\bf k})]}{i(\omega^{(1)}-\omega^{(2)})[1-\exp (-i\omega^{(j)}/T)]} ,
\label{g11}
\end{equation}
\begin{equation} 
g_{12}({\bf k})= \sum_{j=1}^{2} \frac{(-1)^j \Sigma_{12}({\bf k})}{i(\omega^{(2)}-\omega^{(1)})[1-\exp (-i\omega^{(j)}/T)]} .
\label{g12}
\end{equation}

    Thus, we derive the equations for the values of normal and anomalous correlation functions at $(1+p)$ difference position vectors ${\bf l}= {\bf r_l}-{\bf r_0}$ of the neighboring spins:
\begin{equation} 
g_{1j}(l)= \Big( \frac{a}{L} \Big)^d \sum_{\bf k}g_{1j}({\bf k})e^{i{\bf k}{\bf l}}, \ j=1,2; \ l=0,1,..., p.
\label{g1j-consistency}
\end{equation}
Their right hand side is determined by a left hand side $g_{1j}(l)$ itself via Eqs. (\ref{pSelfEnergyIsing}), (\ref{newSigma(l)}), (\ref{omega12})-(\ref{g12}). They constitute an exact closed system of $2(1+p)$ self-consistency equations. Its finding means a solution to the Ising problem in the same sense as finding of a self-consistency equation in the mean-field theory means a solution to a phase transition problem. One can analyze these explicit self-consistency equations by well-known in the mean-field theory analytical and numerical tools. It is relatively simple for 3D Ising model with $\Sigma_{12}  =0$ and zero anomalous correlations, when only $1+p$ self-consistency equations remain. Moreover, in the isotropic case, when the cross-correlations with all $p$ nearest neighbors are the same, the system is reduced to just two equations.

     Note that the closed exact self-consistency equations exist only for unconstrained, auxiliary basis normal and anomalous auto- and cross-correlations. When the latter are found, the actual, observable statistical and thermodynamic quantities can be explicitly expressed in terms of those basis correlations, as is shown in Sec. V for the true, constrained order parameter, correlation functions, and joined statistics of spin-bosons' occupations. Thus, the derived exact general solution for 3D Ising model provides a basis for the calculation of all other statistical and thermodynamic characteristics of critical phenomena.  

\section{Conclusions}

    The presented exact general solution to 3D Ising model is based on the rigorous bosonization of the problem and exact reduction of the many-body Hilbert space for the constrained spin bosons with a subsequent rigorous treatment of all constraints. That approach is opposite to the Onsager's \cite{Onsager}, Polyakov's \cite{Polyakov,Botelho1997} and many other similar approaches, which were aimed at a representation of the model with fermions on a lattice. Once that bosonization was made, finding the exact solution to 3D Ising model was predetermined by perfect algebraic structure and symmetry of the exact solution for the total irreducible self-energy, found in Eqs. (\ref{pSelfEnergyIsing}), (\ref{newSigma(l)}). An amazingly powerful tool for an exact analytical computation of the corresponding nonpolynomial averages with constrained operator  functions is based on the regular method of recurrence equations for partial operator contractions (\ref{spin2-contraction}) \cite{PLA2015,PhysicaScripta2015} and its exact general solution in Eq. (\ref{b-solution}). 

    We stress that the derived self-consistency equations (\ref{g1j-consistency}) are exact equations and, contrary to the mean-field-theory equations, are valid both inside the entire critical region and outside it. They are equations for the $2(1+p)$ normal and anomalous auto- and cross-correlations of neighboring spin bosons, where $p$ is a coordination number of the nearest neighbors for a given site in a lattice. They cannot be reduced to a simpler equation for the order parameter alone, as it was attempted in the original Landau approach. The number of unknown nearest-neighbors' correlations and the respective number of independent self-consistency equations can be less than $2(1+p)$ in the cases of zero anomalous correlations and restricted anisotropy, like a case of an isotropic ferromagnetic phase transition. The obtained exact results for the 3D Ising model are expressed via the solution of these self-consistency equations and are more full than the mean-field or renormalization-group ones \cite{Kadanoff,CritPhen-RG1992}, since they include a fine structure and all details of critical phenomena both in mesoscopic and macroscopic systems.

    Another important point is that the exact self-consistency equations (\ref{g1j-consistency})   contain all effects of spin interaction, including an interaction induced by constraints, since they were derived for the full, constrained and $\hat{\theta}$-cutoff, Hamiltonian (\ref{HIsing}). That makes the present method essentially different from the studies of less constrained and, hence, more tractable models, like spherical and similar models, pioneered by Kac et al. \cite{Kac1952,Baxter1989,CritPhen-RG1992}. 

    Remarkably, the exact self-consistency equation (\ref{g1j-consistency}) is the finite-dimensional nonlinear equation for only $2(1+p)$ numerical parameters, that is not even a functional equation, and has very transparent structure, which involves only elementary functions and standard Fourier transform. This is in contrast with the previous attempts of finding an exact solution to the 3D Ising model, including practically intractable Polyakov's representation in terms of non-interacting fermionic lattice strings \cite{Polyakov,Botelho1997} and exact, nonperturbative renormalization group equations (for a review, see \cite{Vicari2002,Berges2002}). Such renormalization-group equations, including the ones derived by Wilson \cite{Wilson} via an $\varepsilon$-expansion of a space dimensionality $d=4-\varepsilon$ for a $s^4$-model, are aimed to approximate the original microscopic Hamiltonian by an effective Hamiltonian for a large-scale part of a field. These exact equations are very difficult to implement due to their complexity. When Wilson tried to show "that the exact renormalization group equations are not hopelessly intractable functional equations" \cite{Wilson}, he still substituted them with an approximate recursion formula, which is mathematically uncontrollable, but yields the first few orders of $\varepsilon$-expansion. So, one must perform approximations and/or truncations, such as in a well-developed method of an effective average action \cite{Kadanoff,CritPhen-RG1992,Berges2002}. In fact, a number of approximations for the renormalization were introduced, including a neglect method, decimation and other real-space renormalization schemes as well as field-theoretical techniques. However, their rigorous relation to a full exact solution of mesoscopic problem is still not completely understood.

    The obtained exact solution is valid also for 2D Ising model and could be compared against results for the order parameter, correlation  functions, etc., derived from Onsager's or similar 2D solutions. That comparison as well as comparison of the obtained exact results with various Monte Carlo, high-temperature series, finite-size scaling and similar renormalization-group calculations and recent conformal bootstrap studies of 3D Ising model (see, e.g., \cite{Vicari2002,Campostrini1999,Campostrini2002,Deng2003,Yurishchev2004,Anjos2007,Preis2009,Feng2010,Hasenbusch2010,Butera2011,ConformalBootstrap2014} and references therein) require a lot of further work and will be discussed elsewhere.

    At the wings of critical region, the discussed correlation functions describe a critical behavior with critical exponents, different from the ones, predicted by the mean-field theory. Specifically, the difference comes from a different behavior of the exact solution for the self-energies and eigen-energies near a singular point ${\bf k}=0$ of the Fourier sum or integral (in a thermodynamic limit) for the true correlation functions at long range $r \to \infty$,
\begin{equation} 
g'_{1j}({\bf r})= \Big( \frac{a}{L} \Big)^d \sum_{\bf k}g'_{1j}({\bf k})e^{i{\bf k}{\bf r}} \approx \Big( \frac{a}{2\pi} \Big)^d \int g'_{1j}({\bf k})e^{i{\bf k}{\bf r}} d^d {\bf k}.
\label{corr-function}
\end{equation}

    On this basis, one can calculate the long-range asymptotics of the latter integral $\sim r^{2-d-\eta}$, that is a critical exponent $\eta$. Moreover, the exact results for the 3D Ising model, given by Eq. (\ref{g1j-consistency}), go beyond the renormalization group results, describe both the mesoscopic and macroscopic systems, and are valid not only for the asymptotics at the wings of critical region, but also for the critical functions at a central part of critical region. 

     The developed method and the results fully preserve a nonlinear, nonanalytical, critical structure of various statistical and thermodynamic quantities in the critical region, contrary to many other approximate or phenomenological approaches, which usually start from some unjustified, ad hoc assumptions, hypotheses or simplifications in the Hamiltonian or in the description, not fully consistent with an actual critical behavior. The obtained exact solution is based on the regular, quantum-field-theoretical analysis of Green's functions, perfected to an exact rigorous analysis. It yields a systematic, canonical way for a solution of the critical phenomena problem in numerous other models and systems, in particular, for a solution of a long-standing and still open 3D critical phenomena problem. This is in sharp contrast with an approach of the exactly solvable models \cite{Baxter1989,CritPhen-RG1992}, which are solvable only due to their degenerate structure, special symmetries, and low dimensionality and, hence, often provide only a caricature of the actual 3D systems, essentially different from a true picture of typical behavior.

    The presented exact general solution becomes especially simple in a particular case of vanishing anomalous correlations and non-zero normal correlations $g_{1{\bf r}}^{1{\bf r'}} =g_{2{\bf r}}^{2{\bf r'}}$. In this case the exact solution for magnetization 
\begin{equation}
\bar{S}^{'z}_{\bf r} =\frac{\text{per} (2+g_1^{-1})_{\{N-1 \}}}{\text{per} (2+g_1^{-1})} -\frac
{1}{2} 
\label{magnetization-per}
\end{equation}
in Eq. (\ref{spinIsing}) is explicitly given by the permanents of matrices, involving the unconstrained correlation matrix. It is straightforward to obtain similar exact explicit solutions for the correlation functions and other characteristics of critical phenomena in terms of these permanents by means of the presented above method. 

    In the Appendix, we derive an exact general formula for a permanent of circulant matrix, originated from an unpublished Gaudin's note \cite{Mehta1977,Minc1978,Minc1987}, generalize it for the submatrices of circulant matrix in Eq. (\ref{solution}), and discuss their approximations, including the ones yielding the results for "mean-field" and noninteracting approximations. Moreover, the representation in Eq. (\ref{solution}) allows us to avoid calculation of an inverse Fourier transform of correlation matrix in equations like Eq. (\ref{magnetization-per}) since the eigenvalues of circulant correlation matrix are equal to Fourier transforms of the correlation functions, directly given by exact self-consistency Eq. (\ref{g1j-consistency}). This analysis reduces calculation of the circulant-matrix permanent to computing a permanent of degenerate Schur matrices which are well-known in the mathematics of discrete Fourier transform. In particular, the permanent of Schur matrices was studied in \cite{GrahamLehmer1976}. Various techniques for calculation of the permanents and related numerical algorithms are under intensive studies in mathematics since finding the Ryser's \cite{Ryser} and similar algorithms (for a scheme, based on random multipliers and applicable to both determinant and permanent of an arbitrary complex matrix, see \cite{Furer}). This field of computational mathematics is experiencing a fast development after discovery of a FPRAS (fully polynomial randomized approximation scheme) \cite{Jerrum1989} (for a complexity classification of the Ising and other spin models, based on FPRAS, see \cite{Jerrum2015} and references therein). Even a popular software Mathematica, in its late (10.3 or higher) versions, includes now a command "Permanent" that allows one to compute the permanent of any matrix with a dimension less or on the order of 20 as efficient as a usual command "Det" does on calculation of determinants. A numerical analysis of the presented exact solution for the 3D Ising model requires a lot of effort and will be given elsewhere.

    The exact solution includes and clearly shows all complexity of critical phenomena in 3D Ising model as well as relates a problem of its numerical simulation to the $\sharp P$-hard complexity class in computational complexity theory. The latter follows from the result for quasi-cumulants and generating cumulants in Eqs. (\ref{2x2qcumulants}), (\ref{2x2gcumulants}) and even from the result in Eqs. (\ref{rho1per}), (\ref{perP_s})-(\ref{rho'2-0per}), (\ref{magnetization-per}) for vanishing anomalous correlations and non-zero normal correlations $g_{1{\bf r}}^{1{\bf r'}} =g_{2{\bf r}}^{2{\bf r'}}$. Both results yield the joined distributions of spin-bosons' occupations via the permanents. According to a famous Valiant's theorem, the problem of computing the permanent of a matrix is $\sharp P$-hard and provides an example of a problem, where constructing a single solution can be done efficiently, but counting all solutions is hard \cite{Valiant1979,ComputComplexity1994}. It means that a full analysis of 3D Ising model by numerical simulations alone is intractable. That fact stresses an importance of the exact general solution, which unveils a remarkably canonical analytical structure of statistics and thermodynamics of critical phenomena and guides to the adequate approximations and asymptotics for their computation.

    In conclusion, we exactly solve for the constraints, the self-energy, the occupation distributions, and the correlations of spin bosons. It is achieved by formulating a rigorous theory of the constrained spin bosons and by doing the quantum-field-theory calculations for that constrained system at a more fundamental, operator level via the method of the recurrence equations (\ref{spin2-contraction}) for partial operator contractions, which reproduce themselves under a partial contraction operation \cite{PLA2015,PhysicaScripta2015}. The perfectly canonical algebraic structure and explicit form of the obtained analytical results allow one to apply a full power of algebra and matrix calculus as well as direct simulations for a scrutiny of 3D Ising model through the presented exact solution. Its detailed analysis and based on it calculations of various particular statistical and thermodynamic characteristics of critical phenomena for 3D Ising model are coming and could take an effort of many researchers for many years, as it was with analysis of Onsager's exact solution for 2D Ising model \cite{Onsager,Zamolodchikov,Baxter1989,Kadanoff,CritPhen-RG1992,Vicari2002,Percus1971,Montroll1963,Wu1976}.

\section*{Appendix: Calculation of the permanents of circulant matrix and its submatrices via Schur matrices}
     We present the exact formula and asymptotics for the permanents of a circulant matrix and its submatrices. It includes a Gaudin's formula for circulant-matrix permanent as a particular case. For an introduction to a theory of permanents and circulant matrices see \cite{Percus1971,Mehta1977,Minc1978,Minc1987}.   

\subsection{Exact general formula for the permanents of a circulant matrix and its submatrices}

     We consider a $n\times n$ circulant matrix $A$ with the complex entries $A_p^q$ at a row $p$ and  a column $q$ $(p,q=1,...,n)$ and with the eigenvalues $\lambda_1, ..., \lambda_n$. Let $[A]_{\{ i_k, k=1,...,m\}}$ denotes its $(n-m)\times (n-m)$ submatrix, obtained from the matrix $A$ by deletion of the $m$ rows and $m$ columns, which intersect at the diagonal entries $A_{i_k}^{i_k}$, specified by an arbitrary subset of integers $\{ i_k, k=1,...,m \} \subset \{1,...,n \}$; $m$=0,1,...,$n$-1. We prove that its permanent is equal to
\begin{equation}    
\text{per} [A]_{\{ i_k\}} = \frac{1}{n^{n-m}}\sum_{\nu} \frac{\lambda_1^{\nu_1}...\lambda_n^{\nu_n}}{\nu_1!...\nu_n!} |\text{per} S_{\nu \{ i_k \}}|^2 . 
\label{solution}
\end{equation}
Here the sum runs over all $n$-tuples $\nu=(\nu_1,...,\nu_n)$ of nonnegative integers with the sum $\nu_1+...+\nu_n =n-m$. The $S_{\nu \{ i_k \}}$ denotes the $(n-m)\times (n-m)$ submatrix, obtained from a $n\times n$ degenerate Schur matrix 
\begin{equation}
S_{\nu p}^q = e^{\frac{2\pi i pf_{\nu}(q)}{n}}, \  f_{\nu}(q)= 1+ \sum_{t=1}^{n-1} \theta (q-1- \sum_{j=1}^t \nu_j) ,
\label{Snu}
\end{equation}
by deletion of the $m$ rows with indexes $p =i_k, \ k=1,...,m,$ and by truncation of a column-index range to the first $n-m$ values $q=1,...,n-m$. The $\theta (x)$ is the Heaviside step function: $\theta (x)=0$ if $x<0$ and $\theta (x)=1$ if $x \geq 0$. Eq. (\ref{Snu}) means that the first $\nu_1$ columns of the degenerate Schur matrix $S_{\nu}$ are equal to $\theta_n^{(1)}$, the next $\nu_2$ columns are equal to $\theta_n^{(2)}$, and so on, where the $\theta_n^{(t)}$ for $t=1,...,n$ denotes the $t$-th column of the $n\times n$ Schur matrix $\theta_n$, whose $p$-th row and $q$-th column entry is $\exp (\frac{2\pi i pq}{n})$ for $p,q=1,...,n$. The entries of the column $\theta_n^{(t)}$ are equal to $(\theta_n^{(t)})_p = \exp (\frac{2\pi i pt}{n}); \ p=1,...,n$.

     We prove the result in Eq. (\ref{solution}) in detail for the permanent of the full, $n\times n$  circulant matrix $A$, i.e., for the case $m=0$, and then generalize it to the case of an arbitrary $(n-m)\times (n-m)$ submatrix. We start with a well-known representation of the $n\times n$ circulant matrix
\begin{equation}
A_p^q \equiv a_{q-p \ (\text{mod} \ n)} = \frac{1}{n} \sum_{l=1}^n \lambda_l e^{2\pi i (q-p)(l-1)/n} .
\label{circulant}
\end{equation}
It gives the $(p,q)$-th entry of matrix $A$ via its eigenvalues
\begin{equation}
\lambda_l = \sum_{k=1}^n a_{k-1} e^{-2\pi i (k-1)(l-1)/n} , \quad l=1,2,...,n , 
\label{eigenvalues}
\end{equation}
which constitute themselves a discrete Fourier transform of the first row $(a_0, a_1 , ..., a_{n-1})$ of circulant matrix $A$. 

     First, we plug in the representation (\ref{circulant}) into a definition of the permanent $\text{per} A$ and collect terms $\lambda_1^{\nu_1}\lambda_2^{\nu_2}...\lambda_n^{\nu_n}$ with a given $n$-tuple $\nu$ power structure:
\begin{equation}    
\text{per} A = \frac{1}{n^n}\sum_{\nu} \lambda_1^{\nu_1}...\lambda_n^{\nu_n} \sum_{\sigma} \sum_{\{ t_j \}_{\nu}} \prod_{p=1}^n e^{\frac{2\pi i}{n}(\sigma (p)-p)(t_p -1)}. 
\label{per2}
\end{equation}
Here the second sum runs over all permutations $\sigma$ of integers $\{ 1,2,...,n \}$ and the third  sum runs, for a given n-tuple $\nu$, over all sets $\{ t_j \}_{\nu}$ of integers $t_j \in \{ 1,...,n \}$, of which exactly $\nu_j$ integers are equal to $j$ for all $j=1,...,n$, but an order of integers $t_j$ in a set $\{ t_j \}_{\nu}$ is not prescribed. Next, we omit a unity factor $\prod_{p=1}^n \exp[ \frac{2\pi i}{n}(\sigma (p)-p)] =1$ and extend the sum over sets $\{ t_j \}_{\nu}$ to a sum over all permutations $\bar{\sigma}$ of integers $\{ 1,...,n \}$ by replacing $t_p$ with the introduced in Eq. (\ref{Snu}) multi-step function $f_{\nu}(\bar{\sigma} (p))$ and dividing the whole sum by a degeneracy factor $\nu_1!\nu_2!...\nu_n!$, equal to the number of permutations $\bar{\sigma}$ corresponding to the same set $\{ t_j \}_{\nu}$. The result is 
\begin{equation}    
\text{per} A = \frac{1}{n^n}\sum_{\nu} \frac{\lambda_1^{\nu_1}...\lambda_n^{\nu_n}}{\nu_1!...\nu_n!} \sum_{\sigma} \sum_{\bar{\sigma}} \prod_{p=1}^n e^{\frac{2\pi i}{n}(\sigma (p)-p) f_{\nu}(\bar{\sigma} (p))}.  
\label{per3}
\end{equation}

     Second, we use the following factorization 
\begin{equation}    
\prod_{p=1}^n e^{\frac{2\pi i}{n}[\sigma (p) f_{\nu}(\bar{\sigma}(p))-pf_{\nu}(\bar{\sigma} (p))]} = \prod_{p'=1}^n S_{\nu p'}^{\check{\sigma}(p')} \prod_{p=1}^n S_{\nu p}^{*\bar{\sigma}(p)} . 
\label{factorization}
\end{equation}
Here we employ an inverse permutation $\sigma^{-1}$ for a change of variable $p =\sigma^{-1}(p')$  to get the first product as 
\begin{equation}    
\prod_{p'=1}^n S_{\nu p'}^{\check{\sigma}(p')} = \prod_{p=1}^n e^{\frac{2\pi i}{n} \sigma(p) f_{\nu}(\bar{\sigma}(p))} , \ \check{\sigma}(p') = \bar{\sigma}(\sigma^{-1}(p')).
\label{S2}
\end{equation}
The second product employs the complex-conjugated entries $S_{\nu p}^{*q}$ of matrix $S_{\nu}^*$ as follows
\begin{equation}    
\prod_{p=1}^n S_{\nu p}^{*\bar{\sigma}(p)} = \prod_{p=1}^n e^{-\frac{2\pi i}{n}pf_{\nu}(\bar{\sigma}(p))} .
\label{S1}
\end{equation}

     Finally, for each term with a given permutation $\bar{\sigma}$ in Eq. (\ref{per3}), we perform summation over all permutations $\sigma$ via summation of the products in Eq. (\ref{S2}) over permutations $\check{\sigma}$. It yields a common factor $\text{per} \ S_{\nu}$. Then, we perform summation in Eq. (\ref{per3}) over all permutations $\bar{\sigma}$ via summation of the products in Eq. (\ref{S1}). In a results, we factorize the double sum over the permutation sets $\sigma$ and $\bar{\sigma}$ in Eq. (\ref{per3}) into a product of the two single sums over the permutation sets $\check{\sigma}$ and $\bar{\sigma}$, respectively. These single sums are equal to the permanents $\text{per} \ S_{\nu}$ and $\text{per} \ S_{\nu}^*$, respectively. That completes the proof of Eq. (\ref{solution}) for the full circulant matrix when $m=0$. 

    The proof of Eq. (\ref{solution}) for a submatrix with $m \neq 0$ is completely similar to the one presented above. The only difference is that the products in Eqs. (\ref{per2})-(\ref{S1}) include now only $n-m$ factors, instead of $n$ ones in case $m=0$, since all rows with indexes $p =i_k$ and all columns with indexes $q =\sigma (p) =i_k$, $k=1,...,m$, are excluded from the submatrix $[A]_{\{ i_k\}}$ and, hence, from Eqs. (\ref{per2})-(\ref{factorization}). It implies that, instead of the full degenerate Schur matrix $S_{\nu}$, we have now its $(n-m)\times (n-m)$ submatrix $S_{\nu \{ i_k \}}$ both in Eq. (\ref{S1}) (due to exclusion of the rows $p =i_k$) and in Eq. (\ref{S2}) (due to exclusion of the columns $q =\sigma (p) =i_k$ in matrix $A$, which correspond to the rows $p'=\sigma (p) =i_k$ in matrix $S_{\nu}$ because of the change of variable $p =\sigma^{-1}(p')$). Accordingly, all permutations $\sigma , \bar{\sigma}$, and $\check{\sigma}$ in Eqs. (\ref{per2})-(\ref{S1}) are restricted now to the subset of $n-m$ integers $\{1,...,n \} \setminus \{ i_k, k=1,...,m \}$. Also, a prefactor $\frac{1}{n^n}$ in Eqs. (\ref{per2}) and (\ref{per3}) is replaced by a prefactor$\frac{1}{n^{n-m}}$. That completes the proof of the general formula in Eq. (\ref{solution}). In fact, the representation in Eq. (\ref{circulant}) reduces that proof, mainly, to a direct inspection and combinatorics of appropriate terms in the circulant-matrix permanent.

\subsection{Gaudin's formulas for the circulant-matrix permanent and determinant and for the lower and upper bounds of the permanent of a positive semidefinite hermitian circulant matrix}

     A formula for the circulant-matrix permanent
\begin{equation}    
\text{per} A = \frac{1}{n^n}\sum_{\nu} \frac{\lambda_1^{\nu_1}\lambda_2^{\nu_2}...\lambda_n^{\nu_n}}{\nu_1!\nu_2!...\nu_n!} (\text{per} S_{\nu})^2   
\label{per}
\end{equation}
originates from an unpublished Gaudin's note \cite{Mehta1977,Minc1987}. The general result in Eq. (\ref{solution}) immediately yields this Gaudin's formula as its particular case for $m=0$ if one takes into account an equality $\text{per} \ S_{\nu}^* =\text{per} \ S_{\nu}$ valid for any full degenerate Schur matrix $S_{\nu}$.

     For comparison, we present also a similar representation for a determinant of the circulant matrix in Eq. (\ref{circulant}) via the permanents of degenerate Schur matrices:
\begin{equation}    
\text{det} A = \sum_{\nu} \frac{a_0^{\nu_1}a_1^{\nu_2}...a_{n-1}^{\nu_n}}{\nu_1!\nu_2!...\nu_n!} \text{per} S_{\nu} .  
\label{det}
\end{equation}
It follows from (i) a well-known fact that the determinant of a matrix is equal to a product of its eigenvalues, $\text{det} A = \prod_{l=1}^n \lambda_l$, and (ii) a representation of the eigenvalues as the discrete Fourier transform in Eq. (\ref{eigenvalues}), if one employs the n-tuples $\nu$ similar to derivation of Eq. (\ref{per3}). However, a generalization of Eq. (\ref{det}) to a determinant of submatrix $[A]_{\{ i_k\}}$ similar to Eq. (\ref{solution}) is unknown.

     A related Gaudin's formula for the lower and upper bounds of the permanent of a positive semidefinite hermitian circulant matrix, which, therefore, has only nonnegative eigenvalues $\lambda_l \geq 0$, is given in \cite{Mehta1977,Minc1987} as follows
\begin{equation}
\frac{n!}{n^n} \lambda_1^n \leq \text{per} A \leq \frac{n!}{n^n} \xi_n , \quad \xi_n = \sum_{\nu} \lambda_1^{\nu_1}\lambda_2^{\nu_2}...\lambda_n^{\nu_n} ,
\label{per-bounds}
\end{equation}
where $\lambda_1$ is the largest eigenvalue of  $A$ and $\xi_n$ is the coefficient of $z^n$ in the power series expansion of the determinant of resolvent of $A$
\begin{equation}
\frac{1}{\text{det}(1-Az)} = \prod^n_{j=1} (1-\lambda_j z)^{-1} .
\label{inverse det}
\end{equation}

\subsection{Asymptotics for permanents of n$\times$n circulant matrix $A$ and its (n-m)$\times$(n-m) submatrices $[A]_{\{ i_k\}}$}

    If a leading contribution to the circulant-matrix permanent, when the matrix size tends to infinity, is given by a sum in Eq. (\ref{per}) over n-tuples $\nu^{(p)}= (0,...,0,n,0,...,0)$ with only one nonzero component $\nu_p =n$, then one has
\begin{equation}
\text{per} A \to \frac{n!}{n^n} \sum_{p=1}^n \lambda_p^n  \quad \text{at} \quad n \to \infty , 
\label{per-asympt}
\end{equation}
since the permanent of the related degenerate Schur matrix is equal to $\text{per} S_{\nu^{(p)}}=n!$. Note that the permanents $\text{per} S_{\nu}$ for all other n-tuples $\nu \neq \nu^{(p)}$ in Eq. (\ref{per}) are much less than $n!$ and many of them are equal zero exactly. Indeed, they are strongly suppressed due to averaging over the almost homogeneously distributed phases of the terms in each column of the corresponding degenerate Schur matrix $S_{\nu}$. For instance, for any n-tuple $\nu$ with only two nonzero components, one of which is $\nu_p =n-1$ and another one is $\nu_j =1$, the permanent of $S_{\nu}$ is zero: 
\begin{equation}
\text{per} S_{\nu} =0 \ \text{if} \ \nu = \{\nu_p =n-1, \nu_j =1, \nu_k =0 : \ k \neq p,j \} . 
\label{per1+n-1}
\end{equation}
 
     Obviously, the permanent of any $(n-m)\times (n-m)$ submatrix of circulant matrix $A$ has a similar asymptotics
\begin{equation}
\text{per} [A]_{\{ i_k, k=1,...,m\}} \to \frac{(n-m)!}{n^{n-m}} \sum_{p=1}^n \lambda_p^{n-m}  \quad \text{at} \quad n \to \infty 
\label{persubmatrix-asympt}
\end{equation}
if again a contribution from the n-tuples $\nu^{(p)}$ dominates.
 
     In this case, the asymptotics of the ratio of $(n-1)\times (n-1)$-submatrix $[A]_{i}$ and matrix $A$ permanents is 
\begin{equation}
\frac{\text{per} [A]_{i}}{\text{per} A} \to \frac{\sum_{p=1}^n \lambda_p^{n-1}}{\sum_{p=1}^n \lambda_p^n} \qquad \text{at} \quad n \to \infty 
\label{per-ratio}
\end{equation}
for an arbitrary integer $i = 1,..,n$. Here the $\lambda_p$ is the eigenvalue of matrix $A$ and the submatrix $[A]_{i}$ denotes the matrix $A$ with i-th row and i-th column deleted. For $(n-m)\times (n-m)$ submatrix $[A]_{\{ i_k \}}$ with a submatrix-size deficiency $m$, the asymptotics is similar:
\begin{equation}
\frac{\text{per} [A]_{\{i_k, k=1,...,m\}}}{\text{per} A} \to \frac{\sum_{p=1}^n \lambda_p^{n-m}}{\sum_{p=1}^n \lambda_p^n} \quad \text{at} \quad n \to \infty .
\label{per-ratio-m}
\end{equation}
In particular, when there is only one dominant eigenvalue $\lambda_1$, the ratio of submatrix and matrix permanents is
\begin{equation}
\frac{\text{per} [A]_{\{i_k, k=1,...,m\}}}{\text{per} A} = \frac{1}{\lambda_1^{m}} , \quad \text{if} \ \frac{\lambda_l}{\lambda_1} \to 0 \ \ \forall \ l \neq 1. 
\label{dominant-per-ratio-m}
\end{equation} 
 
     Let us consider an example when all eigenvalues are equal to each other, $\lambda_l =\lambda_1, \ l=1,...,n$. It corresponds to the case of noninteracting spins when there are no cross-correlations and the circulant matrix $A$ is equal to its diagonal entry $a_0$ times a unity matrix, $A =a_0 (\delta_{p,q})$, and $\lambda_l =a_0$. In this case the permanents of the matrix $A$ and all its submatrices can be easily calculated directly from the definition of permanent as follows
\begin{equation}
\text{per} [A]_{\{i_k, k=1,...,m\}} = \lambda_1^{n-m} , \quad m=0,1,...,n .
\label{per-diag}
\end{equation}
In this case a formal use of Eqs. (\ref{per-asympt}), (\ref{persubmatrix-asympt}) would greatly underestimate the permanent by a factor $\frac{(n-m)!}{n^{n-m-1}}$, although the ratio of submatrix and matrix permanents 
\begin{equation}
\frac{\text{per} [A]_{\{i_k, k=1,...,m\}}}{\text{per} A} = \frac{1}{\lambda_1^{m}} , \quad \text{if} \ \lambda_l =\lambda_1 \ \forall \ l,
\label{diag-per-ratio-m}
\end{equation}
coincides with the ratio given by Eqs. (\ref{per-ratio}) and (\ref{per-ratio-m}).

     A direct comparison of Eq. (\ref{per-diag}) with the general formula in Eq. (\ref{solution}) yields the following corollary on a sum of squared absolute values of permanents of the submatrices of degenerate Schur matrices over the n-tuples $\nu$:
\begin{equation}    
\sum_{\nu} \frac{1}{\nu_1!...\nu_n!} |\text{per} S_{\nu \{ i_k \}}|^2 = \sum_{\{ t_j \}_{\nu}} |\text{per} S_{\nu \{ i_k \}}|^2 = n^{n-m} , 
\label{sum-schur}
\end{equation}
where the set of degeneracy integers $\{ t_j \}_{\nu}$ is defined after Eq. (\ref{per2}). The result in Eq. (\ref{sum-schur}) constitutes a universal property of degenerate Schur matrices and neither refers nor depends on a form of matrix $A$. It clearly reveals a fact that there is a very large number of various terms $\lambda_1^{\nu_1}...\lambda_n^{\nu_n}$ with nonzero coefficients in front of them under the sum over n-tuples in Eq. (\ref{solution}). These coefficients $|\text{per} S_{\nu \{ i_k \}}|^2 /(\nu_1!...\nu_n!)$ are large enough to make their sum over n-tuples in Eq. (\ref{sum-schur}) approaching infinity as $n^{n-m}$ at $n \to \infty$ for any submatrix-size deficiency $m$.

     Thus, since $(n-m)! \ll n^{n-m}$ at $n-m \gg 1$, the asymptotics in Eqs. (\ref{per-asympt}) and (\ref{persubmatrix-asympt}) or similar ones could be justified only if one or a few eigenvalues strongly dominate all other eigenvalues. Yet, the asymptotics for the ratio of submatrix and matrix permanents in Eqs. (\ref{per-ratio}), (\ref{per-ratio-m}) or similar ones could have a larger region of validity. 
     
     The corollary in Eq. (\ref{sum-schur}) suggests a simplified estimate for the absolute value of an "average" permanent of submatrix $S_{\nu \{ i_k \}}$ of degenerate Schur matrix,
\begin{equation}    
 |\text{per} S_{\nu \{ i_k \}}| \sim \sqrt{(n-m)!} , 
\label{average-per-schur}
\end{equation}       
that is reasonable for a majority of n-tuples $\nu$, if the submatrix $S_{\nu \{ i_k \}}$ includes enough number of entries with different phases for their effective averaging due to summation of various terms in the permanent. (Of course, for some special, strongly degenerate cases the permanent could be essentially different, for example, $ |\text{per} S_{\nu \{ i_k \}}| =(n-m)!$ for the n-tuples $\nu^{(p)}= (0,...,0,n,0,...,0)$ and $\text{per} S_{\nu} =0$ for the n-tuples $\nu$ in Eq. (\ref{per1+n-1}).) Plugging in that estimate into Eq. (\ref{solution}), one has
\begin{equation}    
\text{per} [A]_{\{ i_k\}} = \Big( \frac{1}{n}\sum_{l=1}^n \lambda_l \Big)^{n-m} . 
\label{average-solution}
\end{equation}
This estimate has an equivalent simpler form
\begin{equation}    
\text{per} [A]_{\{ i_k\}} = (a_0)^{n-m} , \qquad a_0 = A_1^1 , 
\label{average-solution2}
\end{equation}
since the sum of all eigenvalues of a matrix is equal to the trace of that matrix, $\sum_{l=1}^n \lambda_l = \text{Tr} A$. Thus, within that "mean-field" approximation, the ratio of submatrix and matrix permanents is given by the following formula
\begin{equation}
\frac{\text{per} [A]_{\{i_k, k=1,...,m\}}}{\text{per} A} = \frac{1}{(a_0)^m} .
\label{mean-field}
\end{equation}
It is different from Eqs. (\ref{dominant-per-ratio-m}) and (\ref{diag-per-ratio-m}), which are valid when either there is one dominant eigenvalue or all eigenvalues are equal to each other.

    Finally, it is worth noting that the presented in this paper exact general solution to 3D Ising model yields the usual approximation of noninteracting spins in external magnetic field, including a  Langevin paramagnetic equation for magnetization, if one employs the approximation in Eq. (\ref{diag-per-ratio-m}) for the related permanents. The mean-field (random-phase) approximation corresponds to the approximation of permanents in Eq. (\ref{mean-field}). 
 
$ $ 

{}
\end{document}